\documentclass[a4paper,pre,twocolumn,amsmath,amssymb,showpacs]{revtex4}

\usepackage{graphicx}

\DeclareMathOperator{\sech}{sech}

\newcommand{\conj}[1]{{#1}^{\ast}}
\newcommand{\order}[1]{{\mathcal O}\left(#1\right)}
\newcommand{\be}{\begin{equation}}
\newcommand{\ee}{\end{equation}}

\begin{document}
\title{Resonantly driven wobbling kinks}
\author{O.F. Oxtoby}
\email{Oliver.Oxtoby@gmail.com}
\affiliation{
CSIR Computational Aerodynamics, Building 12, P.O.~Box 395, Pretoria
0001, South Africa
}
\author{I.V. Barashenkov}
\email{igor@odette.mth.uct.ac.za; Igor.Barashenkov@uct.ac.za}
\affiliation{Department of Mathematics, University 
of Cape Town, Rondebosch 7701, South Africa}
\date{Last update on \today}

\begin{abstract}

The amplitude of oscillations of the freely wobbling kink 
in the $\phi^4$ theory decays due to the emission of 
second-harmonic radiation.
We  study the compensation of these radiation losses (as well
as additional dissipative
losses) 
by the resonant driving of the kink.
 We consider
both  direct and parametric driving at
a range of resonance frequencies.
In each case, we derive 
the amplitude equations which describe the evolution of the 
amplitude of the wobbling and the kink's 
velocity.
These equations predict multistability and  hysteretic transitions 
in the wobbling amplitude for each driving frequency ---
the conclusion verified by numerical simulations of the full
partial differential equation.
We show that the strongest parametric 
resonance occurs when the 
driving frequency 
equals the natural wobbling frequency and not double
that value.  For direct driving, 
the strongest resonance is 
at half the natural frequency, but there is also a weaker 
resonance when the driving frequency equals
 the natural wobbling frequency itself.
We show that this resonance is accompanied by translational motion
of the kink. 
\end{abstract}


\maketitle

\section{Introduction}

The kink of the $\phi^4$ equation
has a mode of internal oscillation, commonly referred to 
as the wobbling mode. To check whether the resonant excitation of 
the wobbling mode
could provide
a channel for pumping energy into a
kink-bearing system, 
 several authors have 
  studied the 
dynamics of  $\phi^4$ kinks  subjected to  resonant direct  
or parametric 
driving and damping
\cite{Gonzalez,QuinteroLetter,QuinteroDirect,QuinteroParametric}.
The damped-driven $\phi^4$ theory arises
 in a variety of physical contexts, in particular
in the description of
  topological-defect dynamics in media with 
temporally \cite{KSV} and
spatially \cite{Konotop}
modulated parameters or in the presence of fluctuations \cite{fluctuations}.
Examples include the drift of domain walls 
in  ferromagnets in oscillatory magnetic
fields  \cite{Sukstanskii};
 the Brownian motion of string-like objects on a 
periodically modulated bistable
substrate \cite{Borromeo};
ratchet dynamics of kinks in a lattice of point-like inhomogeneities
\cite{Molina} and rectification in Josephson junctions
and optical lattices \cite{MQSM}. The damped $\phi^4$ equation 
driven by noise was used to study the 
production of topological defects 
during the symmetry-breaking phase transition
\cite{Zurek}
and spatiotemporal stochastic resonance
in a chain of bistable elements \cite{MGB}.

The mathematical analysis of the damped-driven kinks
 started with the work of
Kivshar, S\'anchez and V\'azquez \cite{KSV} who studied the discrete
parametrically pumped $\phi^4$ system. Assuming that the 
driving frequency lies
above the phonon band
and using the method of averaging, they have discovered
the phenomenon of kink death for sufficiently large driving strengths. 
Next, in an influential paper \cite{Sukstanskii},
Sukstanskii
and Primak  considered the continuous $\phi^4$
equation with a combination of direct and parametric driving.
(See also a related discussion in \cite{discussion}.)
Using a variant of the Lindstedt-Poincar\'e technique where
the velocity of the 
kink is adjusted so as to eliminate secular terms at the lowest
orders of the perturbation expansion in powers of the wobbling
amplitude, they detected a slow unidirectional
motion of the kink.  In their analysis, Sukstanskii
and Primak were not concerned with terms 
higher than quadratic in the amplitude
 of the wobbling; in fact, their approach is not suitable to deal with
 secular terms at the $\epsilon^3$-order of the expansion.
 Neither can it be utilised in the case of the resonant driving
 frequency. 

The resonant situation was studied by
  Quintero, S\'anchez and Mertens 
  \cite{QuinteroLetter,QuinteroDirect,QuinteroParametric}
  who employed the 
method of projections. The method assumes a specific functional 
dependence of the kink on the ``collective co-ordinates", which in 
this case are the width and position of the kink. 
 Inserting the chosen ansatz
  in the partial differential equation
  and projecting the result onto the neutral modes 
  associated with the two degrees
  of freedom,  one obtains a two-dimensional
  dynamical system, a simplification from the 
infinite degrees of freedom present in the 
original partial 
differential equation. 
(In the undamped undriven situation, the collective coordinate
 approach was pioneered 
by Rice and Mele
\cite{RM,Rice}.)
The major advantage of the  method is that if the collective
coordinates have been chosen such that they capture the
essentials of the dynamics, one should be able to uncover the 
very mechanism of the observed nonlinear phenomena --- which
is otherwise concealed by an infinite number of degrees of freedom
(see e.g. \cite{Variational}).
The drawback of the collective coordinate approach, however, is
 that one cannot know beforehand which degrees of 
 freedom are essential and which can be omitted without a qualitative 
 impact on the kink's dynamics.  Specifically, 
 the role of radiation 
 (which is neglected in the approach in question) is not obvious 
 and therefore one has no {\it a priori\/} guarantee
 that a radiationless ansatz is adequate.
 Another disadvantage is that the resulting two-dimensional
 dynamical system is amenable to analytical study 
 only in a very special case (when the damping is set
 to zero \cite{QuinteroLetter,QuinteroDirect,QuinteroParametric}).

In this paper, we approach the resonantly driven wobbling kink from 
a complementary perspective. Instead of trying to guess the most
pertinent set of collective variables, we construct the wobbler as a singular 
perturbation expansion using a sequence of space and time scales.
The nonsecularity conditions yield  
equations for the 
slow-time evolution of the wobbling amplitude and kink velocity.
This asymptotic
procedure has already been implemented
for the unperturbed $\phi^4$ equation \cite{BO};
here we extend it to include damping and driving terms.
We consider both external (direct) and parametric driving,
at several resonant frequencies. This includes the cases 
considered by Quintero, S\'anchez and Mertens in 
\cite{QuinteroLetter,QuinteroDirect,QuinteroParametric}.
Although the multiscale expansion cannot crystallise the 
``minimum set" of degrees of freedom accountable for 
the observed behaviour, it does not suffer from the arbitrariness 
associated with the choice of collective-coordinate ansatz. The multiscale
expansion is asymptotic, i.e. only valid for small amplitudes of
the wobbling mode; however, in the small-amplitude limit the expansion 
provides a faithful description of the wobbler, independent of
any assumptions and mode pre-selections.
Importantly, it 
 does not neglect the radiation.

In all four cases of the direct and parametric driving 
considered in this paper, we derive an autonomous
system of equations for the amplitude of the
wobbling mode and velocity of the kink. 
In each of the four cases this dynamical
system turns out to exhibit stable fixed points
corresponding to the nondecaying wobbling of the kink.
In some parameter regimes, the amplitude
of stable wobbling  is nonunique
and may undergo hysteretic transitions between two nonzero values.
In another case, the wobbling is necessarily accompanied
by translational motion of the kink. 
The conclusions of our
 asymptotic analysis have been verified in direct numerical simulations 
 of the corresponding partial differential equation. 
  The numerical procedure we shall be using throughout 
  this paper was specified in \cite{BO}.

An outline of this  paper is as follows. 
 In sections
\ref{11parametric}  and \ref{21parametric} we study 
the {\it parametrically\/} driven wobbling kink.
In section \ref{11parametric} the frequency of the driver 
is chosen near the natural 
wobbling frequency of the kink
while in \ref{21parametric}, we take
the forcing frequency close to double that value.
In the next section (\ref{h_vs_sh}), we compare the mechanisms that are
at work in each of the two cases.
Subsequently, we
consider the kink driven {\it directly\/} --- first near half of its 
natural wobbling
frequency (section
\ref{12direct}) and then close to the wobbling frequency itself (section \ref{11direct}).
Our conclusions are summarised in section
\ref{Conclusions}.
Here, in particular, we rank the four resonances according
to the amplitude of the resulting stationary wobbling and
according to the width of the resonant frequency range.

\section{$1:1$ parametric resonance}
\label{11parametric}

\subsection{Asymptotic Multiscale Expansion}

We start with the {\it parametric\/} driving
of the form
\begin{equation}
\tfrac{1}{2}\phi_{tt}-\tfrac{1}{2}\phi_{xx} 
+ \gamma \phi_t - \left[1+h\cos(\Omega t)\right]\phi+\phi^3 = 0.
\label{dd}
\end{equation}
This type of a driver was previously considered by 
Quintero, S\'anchez and Mertens \cite{QuinteroParametric}.
The driving frequency $\Omega$ is assumed to be slightly detuned from 
$\omega_0$,
the linear wobbling frequency of the undriven kink: 
\begin{align*}
\Omega = \omega_0(1+\rho).
\end{align*}
We remind the reader that $\omega_0 = \sqrt{3}$ \cite{BO}. 
Introducing a small parameter 
$\epsilon$ (which will be used to measure the amplitude of the wobbling mode
in what follows)
we choose the detuning in the form
\begin{align*}
\rho &= \epsilon^2 R,
\end{align*}
where $R$ is of order one.
Since the frequency of the free 
{\it nonlinear\/} wobbling  is
smaller than $\omega_0$ \{see Eq.(44) in \cite{BO}\},  
we expect that the strongest resonance will occur
not when $\Omega=\omega_0$ but for some small negative $\rho$.
[This will indeed be the case; see Eq.\eqref{rho_res} below.]

Next, the parameters $\gamma>0$ and $h>0$ are the small damping
coefficient  and driving strength, respectively. 
 We choose the following scaling laws for
these parameters:
\begin{equation}
\gamma = \epsilon^2 \Gamma, \quad
h = \epsilon^3 H, 
\label{sca}
\end{equation}
where $\Gamma$ and $H$ are quantities of order $1$. 
This choice of scalings will give rise to amplitude
equations featuring the driving term of the same order of magnitude
as the linear and nonlinear  damping terms
(so that the stationary wobbling regimes become possible). 
We assume that the kink moves with a slowly varying, small velocity:
 $v = \epsilon V$  where
$V = V(T_1, T_2, \ldots)$ is of order 1.

Before embarking on the perturbation expansions, we transform
Eq.\eqref{dd} to the co-moving reference frame:
\begin{align}
\frac12 (1+\rho)^2 \phi_{\tau \tau} - v (1+\rho) \phi_{\xi \tau} 
- \frac{v_{\tau}}{2} (1+\rho) \phi_{\xi} 
- \frac{1-v^2}{2}\phi_{\xi\xi}  \nonumber \\ \label{mofra}
- \phi + \phi^3 = h\cos(\omega_0 \tau) \phi
+\gamma v \phi_\xi 
-\gamma (1+\rho) \phi_{\tau}.
\end{align}
Here
\begin{equation}
\xi = x - \int_0^t v(t') dt'.
\label{xitau}
\end{equation}
We have also changed 
$t \rightarrow \tau$, where
\[\Omega t = \omega_0 \tau.\]

We now expand the field $\phi(\xi, \tau)$ about the kink
$\phi_0 \equiv \tanh \xi$:
\begin{align}
\label{phiexpans}
\phi &= \phi_0 + \epsilon\phi_1 + \epsilon^2\phi_2 + \ldots.
\end{align}
We also define ``slow" space
and time variables
\[
X_n \equiv \epsilon^n \xi, \quad
T_n \equiv \epsilon^n \tau, \quad n=0,1,2,...,
\]
with the standard short-hand notation
$\partial_n  =  {\partial}/{\partial X_n}$, 
$D_n  = {\partial}/{\partial T_n}$.
Substituting \eqref{phiexpans} into the $\phi^4$ equation 
\eqref{mofra}, making use of the chain-rule expansions
${\partial}/{\partial \xi} = 
\partial_0 + \epsilon \partial_1 
+ \epsilon^2 \partial_2 + \ldots$, 
${\partial}/{\partial \tau} = D_0 + \epsilon D_1 
+ \epsilon^2 D_2 + \ldots$,
 and equating coefficients of like powers of $\epsilon$,
we obtain a sequence of linear
partial differential equations --- just as
we have done in the case of the free wobbling \cite{BO}.  
 As in the case of the undamped undriven $\phi^4$ equation, 
 the first-order perturbation is chosen to include only
the wobbling mode,
 \be
 \phi_1= A(X_1, ...; T_1, ...) \sech X_0 \tanh X_0 e^{i \omega_0 T_0} +
 c.c.,
 \label{wobbmo}
 \ee
 while the quadratic correction satisfies the partial differential
 equation
\be
\tfrac{1}{2}D_0^2\phi_2 + {\mathcal L}\phi_2 
= F_2(X_0,...; T_0,...)
\label{qua}
\ee
with 
\[
{\cal L}= -\tfrac12 \partial_0^2 -1+ 3 \phi_0^2
= -\tfrac12 \partial_0^2 +2 - 3 \sech^2 X_0
\]
and
\begin{eqnarray}
F_2= (\partial_0\partial_1 - D_0 D_1)\phi_1 
- 3\phi_0\phi_1^2+ VD_0\partial_0\phi_1 \nonumber \\
+ \frac{1}{2}D_1V\partial_0\phi_0-\frac{1}{2}V^2\partial_0^2\phi_0.
\label{F2a}
\end{eqnarray}
As in \cite{BO}, 
 the second-order perturbation is taken to consist just of the harmonics
present in the forcing function \eqref{F2a}:
 \begin{align}
\phi_2 = \varphi_2^{(0)} + \varphi_2^{(1)}e^{i\omega_0 T_0} + c.c. 
+ \varphi_2^{(2)}e^{2i\omega_0 T_0} + c.c..
\label{phi2m} 
\end{align}
Here  the coefficient functions $\varphi_2^{(0)}(X_0)$, 
$\varphi_2^{(1)}(X_0)$,
and  $\varphi_2^{(2)}(X_0)$ are found by solving
the corresponding ordinary differential equations.
The solvability condition for the first of these equations
is $D_1V=0$ and the solution is
\begin{eqnarray}
\varphi_2^{(0)} = 2 |A|^2\sech^2X_0 \tanh X_0 \nonumber \\
+ \left( \frac{V^2}{2}- 3|A|^2\right) 
 X_0\sech^2X_0 \label{ph20}  \end{eqnarray}
(see \cite{BO}). The solution of the last equation is 
 \be
\varphi_2^{(2)} = A^2 f_1(X_0), \label{ph22}
\ee
with
\begin{multline}
f_1(X_0) = \tfrac{1}{8} \big\{ 6\tanh X_0 \sech^2 X_0 \\
+ (3-\tanh^2 X_0 +ik_0 \tanh X_0) [\conj{J}_2(X_0)-J_2^{\infty}] e^{ik_0 X_0} \\
+ (3-\tanh^2 X_0 -ik_0 \tanh X_0)J_2(X_0) e^{-ik_0X_0} \big\}.
\label{f1}
\end{multline} 
Here the function $J_2(X_0)$ is defined by the integral
\be
\label{Jdef}
J_n(X_0) = \int_{-\infty}^{X_0} e^{ik_0\xi} \sech^n \xi \; d\xi
\quad (k_0=\sqrt{8}),
\ee
with $n=2$. The constant  $J_2^{\infty}$ is the
asymptotic value of $J_2(X_0)$ as $X_0 \to \infty$:
\be
J_n^{\infty}  =\lim_{X_0 \rightarrow \infty}
J_n(X_0).
\ee

Finally, the nonsecularity condition associated with the equation for the coefficient
function $\varphi_2^{(1)}(X_0)$ is $D_1A=0$; with this condition in
place, the solution $\varphi_2^{(1)}(X_0)$ is bounded for all $X_0$
and decays as $|X_0| \to \infty$.
However, this decay is not fast enough \cite{BO}; hence the term $\varphi_2^{(1)}(X_0)e^{i \omega_0 T_0}$
has a {\it quasi}secular behaviour at the infinities and has to be set to zero.
This is achieved by imposing the
condition \cite{BO}
\be
\partial_1 A + i\omega_0 VA = 0. \label{par10}
\ee

 Proceeding to the order $\epsilon^3$, we find
the PDE
\begin{equation}
\label{superharmeps3}
\tfrac{1}{2}D_0^2\phi_3 + {\cal L} \phi_3 =F_3,
\end{equation}
where
\begin{widetext}
\begin{eqnarray}
F_3 = (\partial_0 \partial_1 - D_0 D_1) \phi_2 + 
(\partial_0 \partial_2 - D_0 D_2) \phi_1
+ \tfrac{1}{2}(\partial_1^2 - D_1^2)\phi_1 - \phi_1^3 - 6 \phi_0 \phi_1 \phi_2
+ V D_0 \partial_0 \phi_2 + V D_0 \partial_1 \phi_1 \nonumber \\
 + V D_1 \partial_0 \phi_1
+ \tfrac{1}{2} D_2 V \partial_0 \phi_0 - \tfrac{1}{2} V^2 \partial_0^2 \phi_1
- \Gamma D_0 \phi_1 + \Gamma V \partial_0 \phi_0 - RD_0^2 \phi_1
+ \tfrac{1}{2}He^{i\omega_0 T_0} \phi_0 + c.c.
\label{F3p11}
\end{eqnarray}
\end{widetext}
The cubic correction $\phi_3$ consists of harmonics present 
in the function $F_3$. 
The solvability condition for the first 
harmonic gives the amplitude equation
\begin{equation}
\label{paramomegaamp}
iD_2A + \frac{\omega_0 \zeta}{2}  |A|^2A 
-\omega_0 \left(  R + \frac{V^2}{2} \right) A  
 = \frac{\pi\omega_0}{8}H-i\Gamma A,
\end{equation}
while the solvability condition for the zeroth
harmonic produces 
\[
D_2 V = -2\Gamma V.
\]
Letting $a=\epsilon A$ and keeping in mind that $D_1 A=0$ and $D_1 V=0$,
we can rewrite these two equations in terms of the 
unscaled variables. This gives
a system of two master equations
\begin{subequations}
\label{paramomegaamplitude}
\begin{eqnarray}
{\dot a} = -\gamma a - i \omega_0 \rho a +
 \frac{i}{2} \omega_0 \zeta|a|^2a - \frac{i}{2} \omega_0 v^2a \nonumber
 \\
 - 
 i \frac{\pi}{8} \omega_0 h + \order{|a|^5}, 
 \label{paramomegaamplitudea}
 \\
\label{paramomegaamplitudev}
{\dot v} = -2\gamma v + \order{|a|^5},
\end{eqnarray}
\end{subequations}
where the overdots indicate differentiation with respect to $t$
and the complex coefficient $\zeta$ was evaluated in \cite{BO}:
\begin{equation}
\zeta =\zeta_R+ i \zeta_I= -0.8509+i \, 0.04636. 
\label{zeta}
\end{equation} 

\subsection{Reduced Two-Dimensional Dynamics}

Since $a$ is complex, Eq.\eqref{paramomegaamplitude} defines a dynamical system 
in three dimensions. 
However, equation \eqref{paramomegaamplitudev} will 
damp the variable $v$ until it is of order $|a|^3$
and this will make the term $v^2a$ negligible 
in Eq.\eqref{paramomegaamplitudea}.
Thus, after an initial transient, the dynamics will be 
determined by the two-dimensional system \eqref{paramomegaamplitudea}
with $v=0$. 
Next, the natural wobbling amplitude $a$ may depend,
parametrically,  on $\xi$.
However, equation \eqref{par10} and the fact that $v \to 0$ as $t$ grows, 
imply that $a$ may only depend on $\xi$ via $X_2$, $X_3$, etc. That is,
the dependence is weak.

Letting $a=re^{-i \theta}$, Eq.\eqref{paramomegaamplitudea} yields
\begin{equation}
{\dot r}= - \frac{\omega_0}{2} \zeta_I r^3 + \gamma (r_0 \sin \theta -
r),
\label{dec}
\end{equation}
where $r_0=\frac{\pi}{8} \omega_0 h / \gamma$. For all $a$ with $|a|>r_0$,
the right-hand side of \eqref{dec} is negative and so 
no trajectories can escape to infinity. 
On the other hand, applying Dulac's criterion
(with Dulac's function equal to a constant), one can easily ascertain that 
Eq.\eqref{paramomegaamplitudea} 
with $v = 0$  does not have closed orbits. 
Hence, all trajectories must
flow towards fixed points with finite $|a|$.

The fixed points of the system \eqref{paramomegaamplitudea}
are given by the equation
\begin{equation}
 (\gamma  + i\omega_0 \rho) a -
 \frac{i \omega_0 \zeta}{2}  |a|^2a =
 -\frac{\pi i \omega_0}{8}h.
 \label{pi8}
 \end{equation} 
From Eq.\eqref{pi8}, the absolute value of $a$ satisfies 
  \begin{equation}
  {\cal H}(|a|^2)=h, 
  \label{modu}
  \end{equation}
  where the  function ${\cal H} (|a|^2)$ is defined by
  \begin{equation}
  {\cal H}^2=
  \frac{64}{\pi^2} |a|^2 \left[
  \left(\frac{\gamma}{\omega_0} + \frac{ \zeta_I}{2} 
  |a|^2\right)^2
  + \left(\rho -\frac{\zeta_R}{2} 
   |a|^2\right)^2\right].
  \label{cubic}
 \end{equation}

Assume, first, that $\rho>\rho_0$, where 
\begin{equation}
\rho_0(\gamma)=  \frac{1}{\omega_0}\frac{ \zeta_I -\sqrt{3} \zeta_R}{ \zeta_R + \sqrt{3} \zeta_I} \gamma
 = -1.139 \gamma.
\label{rhoga}
\end{equation}
In this case
${\cal H}(|a|^2)$ 
  is a monotonically growing function, with the range
$(0, \infty)$.
Eq.\eqref{modu} has a single positive root $|a|^2$ for any $h$
and the dynamical system \eqref{paramomegaamplitude}
has a single stationary point. This fixed point is always stable. 

Now let  $\rho< \rho_0$. Here, the range of the function
${\cal H}(|a|^2)$ is still $(0,\infty)$; however the
function  grows for
small and large values of $|a|^2$ but  decreases in the
intermediate interval  $|a_-|^2< |a|^2 < |a_+|^2$,
where
\begin{subequations}\label{apm}
\begin{eqnarray} 
|a_{\pm}|^2= \frac23 \frac{2 \alpha \pm \sqrt{\alpha^2- 3
\beta^2}}{|\zeta|^2}, \\
\alpha= \zeta_R \, \rho - \zeta_I \, \frac{\gamma}{\omega_0};
\quad
\beta= \zeta_I \, \rho + \zeta_R \, \frac{\gamma}{\omega_0}.
\label{alpha_beta}
\end{eqnarray}
\end{subequations}
Consequently, Eq.\eqref{modu} has a single root for small and large
values of $h$ and three roots in the intermediate region defined by
$ h_+<h < h_-$,
where 
\begin{equation}
h_\pm = {\cal H}(|a_\pm|^2).
\label{pmH}
\end{equation}
 For the dynamical system 
\eqref{paramomegaamplitude} this implies that 
there is only one fixed point (which is stable) for
small and large $h$, but as 
 $h$ approaches the value $h_+$ from below
or the value $h_-$ from above, two new fixed
points are born in a saddle-node bifurcation. The region $ h_+<h < h_-$
is characterised by bistability; an adiabatic variation of
$h$ will result in hysteretic transitions between two stable fixed
points.

\begin{figure}
\includegraphics{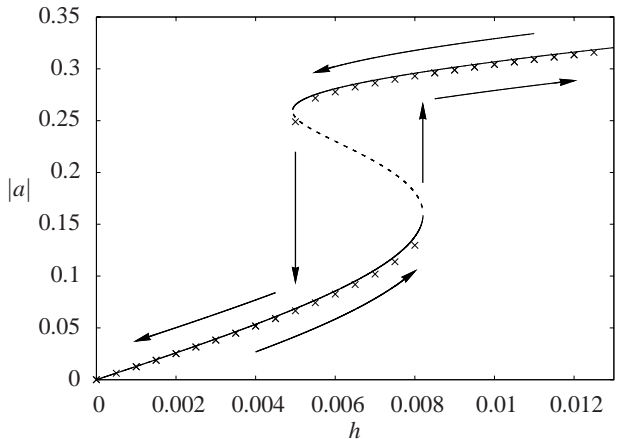}
\caption{\label{hyster} The hysteresis loop
observed in the $1:1$ parametrically driven 
$\phi^4$ equation, Eq.\eqref{dd},
with $\gamma=0.01$ and $\rho=-0.03$. 
The driving strength $h$ is increased from
$0$ to $0.0125$ in increments of $5 \times
10^{-4}$, and then reduced  back to $0$
as indicated by arrows.
Crosses mark simulations of the equation \eqref{dd}; continuous
and dashed 
 lines depict stable and 
 unstable fixed points of the amplitude
equation \eqref{paramomegaamplitudea} with $v=0$.
}
\end{figure}

The existence of hysteresis has been verified in the 
direct numerical simulations of Eq.\eqref{dd}, with $\gamma=0.01$.
Starting with $h=0$ we increased $h$ 
past $h_-$, and then reduced it back to zero. At each $h$-step,
we used the final values of $\phi(x)$ and $\phi_t(x)$ from
the previous-step simulation as initial conditions
for the new run.
For each $h$
we measured the value of $a$ to which the numerical 
solution settled after transients  died out. 
The resulting amplitude $|a|$ is shown in Fig.\ref{hyster}; clearly visible is 
the hysteresis loop.
Fig.\ref{hyster} corresponds to simulations with $\rho=-0.03$; 
for $\rho=-0.02$ the hysteresis loop is smaller and for 
$\rho=-0.01$ it disappears completely. This is consistent with
the value of $\rho_0$ given by Eq.\eqref{rhoga}.
The value $h_-=0.008$ at which the amplitude was 
recorded to jump from the bottom
to the top branch in Fig.\ref{hyster}, and the value $h_+=0.005$
at which it dropped back as $h$ was decreased, are
also in agreement with the corresponding predictions
of the amplitude equation. Namely, Eqs.\eqref{apm},\eqref{cubic}
and \eqref{pmH} 
give $h_-= 8.2 \times
10^{-3}$ and $h_+ =4.9 \times 10^{-3}$.

For any given $h$, equation \eqref{modu}
can be regarded as a quadratic equation for the detuning $\rho$
where the coefficients are explicit functions
of $|a|^2$. 
 There are two roots $\rho_{1,2}$ for $|a|^2$
smaller than $|a_{\rm res}|^2$, and none for $|a|^2 > |a_{\rm res}|^2$, 
where $|a_{\rm res}|^2$ is a unique positive root of the 
equation
\begin{equation}
\frac{64}{\pi^2}|a_{\rm res}|^2
\left( \frac{\gamma}{\omega_0}+ \frac{\zeta_I}{2} |a_{\rm res}|^2
\right)^2
=  h^2.
\label{maxa}
\end{equation}
The value $|a_{\rm res}|$ defined by \eqref{maxa} gives the 
largest amplitude of the wobbling achievable for the 
given driving strength $h$. The corresponding value
of the detuning,
\begin{equation}
\rho_{\rm res} = \frac{\zeta_R}{2} |a_{\rm res}|^2<0,
\label{rho_res}
\end{equation}
ensures the strongest resonance. As was expected,
the strongest resonance is achieved with negative
detuning.

\subsection{1:1 Parametrically Driven Wobbler}

For large times, 
the asymptotic expansion for the damped-driven wobbler is given
by Eq.(48) in \cite{BO}                               
where we just need to set $v=0$
and replace $\omega_0$ with $\Omega$:
\begin{multline}
 \phi(x,t) = \tanh 
 \left[
 (1-3|a|^2) \xi \right]
  + a \sech \xi \tanh \xi e^{i\Omega t} + c.c. \\
+ 2 |a|^2 \sech^2 \xi \tanh \xi   
+ a^2 f_1(\xi)e^{2i\Omega t} + c.c. + \order{|a|^3}.
\label{paraexp}
\end{multline}
Here 
$\xi=x-x_0$, where $x_0$ is a constant determined
by initial conditions, and
$a$ is a stable fixed point of the dynamical system
\eqref{paramomegaamplitudea} with $v=0$
[a unique fixed point or one of the two stable fixed points depending 
on whether $h$ is outside or inside the 
bistability interval $(h_+,h_-)$].
 The function $f_1(\xi)$ is given by Eq.\eqref{f1}. 
The interpretation of different terms in \eqref{paraexp}
is the same as in the case of the freely wobbling kink \cite{BO}.

Like the corresponding formula for the free wobbler, 
the expansion 
\eqref{paraexp} 
 is only valid at distances
$|\xi|=\order{1}$.
For larger distances one has to use the outer expansions 
\begin{equation}
\phi=\pm 1 + \epsilon^2 \phi_2+ \epsilon^3 \phi_3+...,
\label{outexp}
\end{equation} 
 with coefficients $\phi_n$ determined as in section V          
 of \cite{BO}. The analysis of the outer equations 
produces results equivalent to those in \cite{BO}:
The second-harmonic radiation propagates away from the core of the kink 
at the group velocity, leaving in its wake 
a sinusoidal wave with the frequency $2 \Omega$,
wavenumber $k_0=\sqrt{8}$ and constant amplitude of the order $|a|^2$.

Unlike the case of  the free
wobbling of the kink,  the frequency of the oscillation
is not determined by  its amplitude but is locked to the 
frequency of the driver, $\Omega$.
  Another difference from the undamped
  undriven case  is that the driven oscillations
  of the wobbler do not die out
  as $t \to \infty$. Instead, the amplitude of the oscillations
 approaches a nonzero constant value which is determined by the 
 parameters of the damping and driving
 and --- in the bistable region --- by the initial conditions.
 On the other hand, the asymptotic velocity of the damped-driven wobbler 
 {\it is\/} zero.

 It is interesting to note that unlike 
 in the case of the parametrically
 driven damped
 linear oscillator \cite{Grimshaw}
 or damped-driven breather of the 
 sine-Gordon or $\phi^4$ equation \cite{BBK},
  there is no threshold driving 
 strength in the case of the damped-driven wobbler. No matter how small is $h$, 
 the amplitude $a(t)$ will not decay to zero
 as $t \to \infty$.

\section{$2:1$ parametric resonance}
\label{21parametric}

\subsection{Asymptotic Expansion}

It is a textbook fact that the strongest parametric resonance is
achieved when the parameter of the oscillator is varied at {\it double\/} 
its natural frequency. With an eye to the detection of
 the most efficient
driving regime for the wobbling kink,  we now consider the driving frequency close to 
twice its natural wobbling frequency:
\begin{align}
\tfrac{1}{2}\phi_{tt}-\tfrac{1}{2}\phi_{xx} 
+ \gamma \phi_t - \left[1+h\cos(2\Omega t)\right]\phi+\phi^3 &= 0.
\label{pd21}
\end{align}
As before,
\begin{align*}
\Omega = \omega_0(1+\rho),\quad \rho=\epsilon^2 R,
 \quad \gamma= \epsilon^2 \Gamma,
\end{align*}
but now we use a different scaling for $h$:
\begin{align*}
h = \epsilon^2H.
\end{align*}
We transform the equation in exactly the same way as 
we did in the previous section; this yields
\begin{multline*}
\frac12 (1+\rho)^2 \phi_{\tau\tau} - v(1+ \rho) \phi_{\xi\tau}
- \frac{v_{\tau}}{2}(1+ \rho) \phi_{\xi} - \frac{1-v^2}{2} \phi_{\xi\xi} \\
-\phi+\phi^3 
= h \cos(2 \omega_0 \tau) \phi + \gamma v \phi_{\xi} 
-\gamma (1+\rho)  \phi_{\tau}.
\end{multline*}

The perturbation expansion is unchanged from the undamped undriven case at 
$\order{\epsilon^1}$.
With the addition of the
$\epsilon^2$-strong driving, the equation at $\order{\epsilon^2}$
acquires
additional terms on the right hand side as compared to 
Eq.\eqref{qua}: 
\begin{align*}
\frac{1}{2}D_0^2\phi_2 + {\mathcal L}\phi_2 &= 
F_2(X_0,...;T_0,...)
 + \frac{H}{2}\phi_0e^{2i\omega_0 T_0} + c.c.
\end{align*}
Here $F_2$ is as in Eq.\eqref{F2a}. 
The zeroth- and first-harmonic components of $\phi_2$ are not
affected by this extra term.
Namely, assuming that the solution is in the form \eqref{phi2m} 
and setting $D_1 V=0$, we get
Eq.\eqref{ph20}           
for  $\varphi_2^{(0)}$ while  
imposing $D_1A=0$,
Eq.\eqref{par10} produces $\varphi_2^{(1)}=0$. As for the coefficient
function $\varphi_2^{(2)}$,
 we obtain
\begin{equation}
\varphi_2^{(2)} = A^2 f_1(X_0)+H f_2(X_0),
\label{p2H}
\end{equation}
where the function $f_2(X_0)$ satisfies 
\begin{align}
\label{radiation2}
({\mathcal L}-6)f_2(X_0) = \frac{1}{2}\tanh X_0.
\end{align}
We note that the value $6$ lies in the continuous spectrum
of the operator $\cal L$, and 
so in order to determine $f_2(X_0)$ uniquely, one has to impose two 
additional conditions fixing the coefficients of two
bounded homogeneous solutions that can be added to $f_2$. 
 We do this by requiring the absence of incoming radiation.
The particular solution of \eqref{radiation2}
which obeys these radiation boundary conditions is
\begin{equation}
f_2(X_0) = -\frac{1}{12} f_1(X_0)+ \frac{1}{24} \tanh X_0 (2 \sech^2 X_0-3),
\label{f2}
\end{equation}
where the function $f_1(X_0)$ is given by  Eq.\eqref{f1}.
 In what follows, we will use the fact
that
$f_2(X_0)$ is an odd function.

The first term in 
the right-hand side of 
\eqref{p2H} describes the familiar second-harmonic radiation 
from the freely wobbling kink. The second term consists of 
the induced second-harmonic radiation and
a standing wave   --- also excited by the  forcing.

At the order $\epsilon^3$, we get the equation \eqref{superharmeps3}
where $F_3$ is given by Eq.\eqref{F3p11} with the term
$\frac12 H e^{i \omega_0 T_0} \phi_0$ replaced
with $\frac12 H e^{2i\omega_0 T_0} \phi_1$. 
The amplitude equation for $A$, 
which arises as the solvability condition for the
first harmonic, is now
\begin{equation}
\label{param2omegaamp}
i D_2A+ \frac{\omega_0  \zeta}{2} |A|^2A - \omega_0 \left( \frac{V^2}{2} 
+ R \right) A = 
 \frac{\omega_0  \sigma}{2}  H \conj{A}-i \Gamma A,
\end{equation}
where
\begin{eqnarray*}
 \sigma =\int_{-\infty}^{\infty} \left[ 
\tfrac{1}{2}\sech^2 X_0 \tanh^2 X_0 \right.
  \\
-\left. 6\sech^2 X_0 \tanh^3 X_0 f_2(X_0) \right] dX_0.
\end{eqnarray*}
The imaginary part of this integral is
\[
\sigma_I = \frac{1}{12} \zeta_I=0.003863;
\]
for the real part we find, numerically,
\[
\sigma_R=0.5958.
\]

The solvability condition for the 0th harmonic gives
\begin{equation*}
D_2 V = -2 \Gamma V.
\end{equation*}

We finally write the amplitude and velocity equations 
in terms of the natural variables
$a=\epsilon A$ and $v=\epsilon V$  and the unscaled time $t$ (as in the
previous section):
\begin{subequations}
\label{param2omegaamplitude}
\begin{align}
\label{param2omegaamplitudea}
\begin{split}
{\dot a} &= -\gamma a - i\omega_0 \rho a + \frac{1}{2}i\omega_0 \zeta |a|^2a \\
&\;\;\; - \frac{1}{2}i\omega_0 v^2a - \frac{1}{2}i\omega_0 \sigma h\conj{a} +
\order{|a|^5},
\end{split}\\
\label{param2omegaamplitudev}
{\dot v} &= -2\gamma v + \order{|a|^5}.
\end{align}
\end{subequations}

\subsection{Reduced Dynamics in Two Dimensions}

As in the previous case of the $1:1$ parametrically
driven wobbler, the velocity tends to zero as $t \to \infty$ 
and the evolution of $a(t)$ is governed by the dynamical
system  \eqref{param2omegaamplitudea} with $v=0$.
This two-dimensional dynamical system does not have periodic orbits, as one can 
readily check using Dulac's criterion.
Letting $a=r e^{- i \theta}$ and $\sigma =|\sigma| e^{i {\rm Arg} \sigma}$,
 Eq.\eqref{param2omegaamplitudea}
yields
\begin{equation}
{\dot r}= - \gamma r +\frac{\omega_0}{2} \zeta_I r 
\left[ r_0^2 \sin( 2\theta + {\rm Arg} \, \sigma)-r^2 \right],
\label{cha}
\end{equation}
where $r_0^2= (|\sigma|/ \zeta_I)h$. Since the right-hand side of 
\eqref{cha} is negative for all $a$ with $|a|>r_0$,
no trajectories can escape to infinity. 
Therefore, all trajectories should flow to one of the fixed points.
  The fixed points 
are given by the equation
\begin{equation}
\gamma a + i \omega_0 \rho a -
i\frac{\omega_0 \zeta}{2}  |a|^2 a =
-i \frac{\omega_0 \sigma}{2}  h \conj{a}.
\label{aste}
\end{equation}
One fixed point is trivial, $a=0$; this fixed point 
 is stable if $h<h_+$,
where
\begin{equation}
\frac{|\sigma|^2}{4} h_+^2=\left( \frac{\gamma}{\omega_0}
\right)^2 + \rho^2,
\label{h_plus}
\end{equation}
and unstable otherwise.
For the nontrivial points, we get
\begin{equation}
{\cal H}(|a|^2)= h,
\label{para1}
\end{equation}
where
\begin{equation*}
{\cal H}^2=  \frac{4}{|\sigma|^2} 
\left[ \left(\frac{\gamma}{\omega_0}
+ \frac{\zeta_I}{2}  |a|^2\right)^2 +
\left( \rho -\frac{\zeta_R}{2}  |a|^2\right)^2 \right].
\label{para2}
\end{equation*}

Assume, first, that $\rho> \rho_0$, where
\begin{equation}
\rho_0(\gamma) = \frac{1}{\omega_0}  \frac{ \zeta_I}{\zeta_R} \gamma
=-0.03146 \gamma.
\label{cri_rho}
\end{equation}
The function ${\cal H}(|a|^2)$
with $\rho$ in this parameter range is monotonically growing,
from $h_+$ to infinity.
The equation \eqref{para1} has  one root provided $h>h_+$,
and no roots otherwise.
Consequently,  in the region $h>h_+$ the
dynamical
system \eqref{param2omegaamplitude}
has 2 stable fixed points $a_1$ and $-a_1$, where
\begin{equation}
|a_1|^2=  \frac{2\alpha+ \sqrt{|\zeta \sigma|^2 h^2-
4\beta^2}}{|\zeta|^2}, 
\label{a1}
\end{equation} with $\alpha$ and $\beta$ as in \eqref{alpha_beta}.
In the region $h<h_+$, the (stable) fixed point at the origin
is the only fixed point available in the system. 
(Thus we have a supercritical pitchfork
bifurcation as $h$ is increased through $h=h_+$.)

Assume now $\rho< \rho_0$. As $|a|^2$ grows from
zero to infinity,
the function $\cal H$ decreases
from $h_+$ to its lowest value
of $h_-$, where 
\begin{equation}
\frac{|\sigma|^2}{4} h_-^2= \frac{1}{|\zeta|^2} \left(\zeta_R
\frac{\gamma}{\omega_0}  +\zeta_I \,\rho  \right)^2,
\label{sih}
\end{equation}
and then increases to infinity.
Therefore, for  $\rho<\rho_0$, the
dynamical
system \eqref{param2omegaamplitude}
has one fixed point at the origin for small driving
strengths $0<h<h_-$;
5 fixed points $a_1, a_2, 0$,  $-a_1$, and $-a_2$
for the intermediate  strengths $h_-<h<h_+$; and
3 fixed points $a_1,  0$, and $-a_1$ for   $h>h_+$.
Here $h_+$ is given by Eq.\eqref{h_plus} and $h_-$ by \eqref{sih}.
The nontrivial fixed points $a_1$ and $a_2$ are 
born in a saddle-node bifurcation at $h=h_-$.
At $h=h_+$, a subcritical 
pitchfork bifurcation occurs; here, the point $a_2$ merges with the 
trivial fixed point. Therefore, out of the two nontrivial fixed points
$a_1$ and $a_2$, the stable one is $a_1$, i.e. the fixed point with
the larger absolute value --- given by Eq.\eqref{a1}.
In summary, for $h<h_-$ all trajectories flow 
to the origin; for $h>h_+$ they are attracted to the
nontrivial fixed points $\pm a_1$, and, finally, in the
region $h_-< h<h_+$ we have a tristability between 
$a=0$ and $a= \pm a_1$.

\begin{figure}
\includegraphics{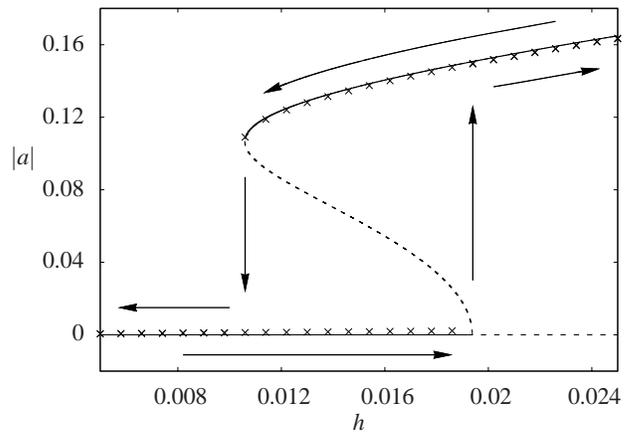}
\caption{\label{simulations21} The hysteresis loop in the
$2:1$ parametrically driven $\phi^4$ equation, Eq.\eqref{pd21},
with $\gamma=0.005$ and $\rho=-0.005$. 
The driving strength is increased from $0.005$ 
to $0.025$ in increments of $8 \times 10^{-4}$, and then reduced back
to $0.005$.
Crosses mark results of simulations of 
Eq.\eqref{pd21}. 
The continuous and dashed lines show the stable
and unstable fixed points of the amplitude equation 
\eqref{param2omegaamplitudea} with $v=0$. }
\end{figure}

These predictions of the amplitude equation were compared
to results of direct numerical simulations of Eq.\eqref{pd21},
with $\gamma=0.005$. As in the previous section, we increased 
$h$ past $h_+$ and then reduced it to values under $h_-$. Fig.\ref{simulations21}
shows the hysteresis loop arising for 
$\rho=-0.005$. For $\rho=-0.003$ the hysteresis was
less pronounced and 
for $\rho=0$ it was seen to disappear completely. These observations
are consistent with the value of the critical detuning 
\eqref{cri_rho} which, for $\gamma=0.005$, equals
$\rho_0=-1.573 \times 10^{-4}$.
The bifurcation values $h_{\pm}$ observed in simulations
with $\rho=-0.005$
($h_+ = 0.019$ and $h_-=0.011$),
are also in agreement with the predictions of the 
amplitude equation (which gives $h_+=0.01938$ and $h_-=0.01059$).

\subsection{2:1 Parametrically Driven Wobbler}

Restricting ourselves to the $t \to \infty$ asymptotic behavior
of $\phi(x,t)$, the leading orders of the  perturbation expansion
in the case of the subharmonic response
are
\begin{multline}
\phi(x,t) = \tanh \left[ (1-3|a|^2) \xi \right]
 + a \sech \xi \tanh \xi  e^{i\Omega t} + c.c. \\
+ 2 |a|^2\sech^2 \xi \tanh \xi  \\
+ \left[a^2 f_1(\xi)+hf_2(\xi)\right]e^{2i\Omega t} + c.c. +
\order{|a|^3}. 
\label{phxt11}
\end{multline}
Here $\xi=x-x_0$;
$a$ is a stable fixed point (zero or nonzero) given by one of 
the roots of \eqref{aste},  and
the functions $f_1$ and $f_2$ are defined by 
Eqs.\eqref{f1} and \eqref{f2}, respectively. 
The main
difference from the case of the $1:1$ parametric resonance is that the
amplitude of the wobbling 
approaches a nonzero  value only if the driver's strength
exceeds a certain
threshold; this threshold value is given by
$h_+$ in the region $\rho> \rho_0$ and by $h_-$ in the region $\rho<\rho_0$.
 If $h$ lies below the threshold, 
the wobbling dies out and we need to set $a=0$ in Eq.\eqref{phxt11}. 
We also note that the $2:1$ resonant driving excites a
standing wave and
radiation with the frequency $2 \Omega$ and amplitude proportional
to
$h$ [the $f_2$-term in \eqref{phxt11}].

As Eq.\eqref{paraexp} of  the previous section, the 
expansion  \eqref{phxt11} is only valid 
on the lengthscale $\xi=\order{1}$. To describe the waveform 
at longer distances, we need to invoke the outer expansions
\eqref{outexp}. Evaluating the coefficient of the term $\epsilon^2$
in these expansions and matching it to the ``inner" expression
\eqref{p2H} in the overlap region, we obtain
\begin{equation}
\phi_2= \pm {\cal J} B_{\pm} 
e^{i(2\omega_{0}T_0 \mp k_{0} X_0)} + c.c.- \frac{H}{4} \cos (2 \omega_0
T_0),
\label{pphi2}
\end{equation}
where the 
top and bottom sign pertain to the regions $X_0>0$ and $X_0<0$,
respectively. In Eq.\eqref{pphi2}, ${\cal J}=(2-i k_0) J_2^\infty$
and the functions $B_\pm = B_\pm(X_1,X_2,...;T_1,T_2,...)$
satisfy  
\begin{equation}
B_\pm (0,0,...;T_1,T_2,...)= A^2(0,0,...;T_2,T_3,...)-
\frac{H}{96}.
\label{hash}
\end{equation}
Eq.\eqref{hash} represents the boundary condition for the 
amplitudes $B_\pm$; the equations of motion for these variables 
arise at the order $\epsilon^3$ and coincide with 
Eqs.(33) of \cite{BO}.                        
The solution of these equations with the boundary condition
\eqref{hash} is qualitatively similar to the solution with the ``undriven"
boundary condition $B_\pm(0,0,...;T_1,T_2,...)=
A^2(0,0,...;T_2,T_3,...)$.                       
Namely, we have two outward-propagating 
waves leaving the amplitudes $B_\pm$ equal to the constant $A^2-H/96$
in their wake. 

\section{Harmonic vs Subharmonic Parametric Resonance: Qualitative Comparison}
\label{h_vs_sh}

With the amount of detail that we had to provide to justify our 
conclusions and derivations, the resonance mechanisms of the 
driven wobbling kink  may not be 
easy to crystallise. The purpose of this short section is to 
discuss the two parametric resonances  
qualitatively --- in particular, to comment on their atypical hierarchy. 

We observed that 
the amplitude $|a|$
is of the order  $h^{1/3}$ 
in the case of  
the harmonic resonance (i.e. the resonance
arising when the driving frequency 
$\omega_{\rm d}$ is 
near $\omega_0$) but only  
$\order{h^{1/2}}$ in the case of
 the subharmonic resonance (the resonance arising for
$\omega_{\rm d} \approx 2\omega_0$). Thus the harmonic resonance is 
stronger than the subharmonic one, and  
this is precisely the opposite behaviour to what we might na\"\i vely 
expect based on our intuition about the parametric driving. 

To explain this surprising behaviour qualitatively, we
write the term $h  \cos(n\Omega t)\phi$ in Eqs.\eqref{dd}
and \eqref{pd21} as
$h \cos (n \Omega t) \phi_0$ plus terms of order $ha$ and smaller.
This representation reveals that what was introduced  as a parametric
driver is, to the leading order, an external (direct) driving force.
This driving force is nonhomogeneous, i.e. its magnitude and direction
vary with the distance, and it has  
odd spatial parity. 
When $n=1$, the frequency of this driving force coincides with the 
natural frequency of the wobbler and its spatial parity  coincides with 
the parity of the wobbling mode (which is also odd). As a result, we
have a strong direct resonance.

When $n=2$, the external force $h  \cos (2 \Omega t) \phi_0$ 
is not in resonance with the wobbling frequency.
However the function $h  \cos(2\Omega t)$ acts as a parametric driver
on the {\it next\/} term in the expansion of $\phi$
--- that is, the product $h  \cos(2\Omega t)\epsilon \phi_1$ has 
 the resonant frequency.
Importantly, this term has the ``correct", odd, parity as a
function of $\xi$.

In addition, 
 the odd-parity
force $h  \cos (2 \Omega t) \phi_0$ generates  odd-parity radiation 
and the odd-parity standing wave, both with the 
frequency $2 \Omega$. This radiation and  standing wave also couple
to the wobbling mode, via the term $\epsilon^3 \phi_0 \phi_1 \phi_2$
in Eq.\eqref{pd21}. This constitues a concurrent
driving mechanism. Since each of the two  mechanisms  
  is indirect
 (i.e. requires the wobbling mode as a mediator
 for the frequency halving)
 and since the resulting effective driving strength is proportional to the 
 amplitude of the wobbling mode (assumed small), 
the response to the frequency $2\Omega$ is weaker than to $\Omega$.

\section{$1:2$ direct resonance}
\label{12direct}

\subsection{Perturbation Expansion}

We start with  the direct driving at \emph{half} the natural
wobbling frequency.
This and the following case of the $1:1$ direct resonance were previously
considered by Quintero, S\'anchez and Mertens \cite{QuinteroLetter,QuinteroDirect} 
and so we will be able to compare our results to theirs.
The equation is
\begin{equation}
\frac{1}{2}\phi_{tt}-\frac{1}{2}\phi_{xx} + \gamma \phi_t - \phi + \phi^3 = 
h\cos \left(\frac{\Omega}{2} t \right),
\label{dd12}
\end{equation}
where, as in the previous sections, $\Omega = \omega_0(1 + \rho)$.
 As before,
 we change the time variable so that $\Omega t = \omega_0 \tau$ and 
transform the equation to the moving frame:
\begin{multline}
\tfrac{1}{2} (1+\rho)^2 \phi_{\tau\tau} - v(1+\rho)\phi_{\xi\tau} 
- \frac{v_{\tau}}{2}(1+ \rho) \phi_{\xi} - \frac{1-v^2}{2}\phi_{\xi\xi} \\
- \phi + \phi^3 = h\cos\left(\frac{\omega_0}{2}\tau\right)
+ v \gamma \phi_{\xi}-\gamma (1+\rho) \phi_{\tau}.
\label{12direq}
\end{multline}

Keeping our standard scalings
for the small parameters 
$\gamma$ and $\rho$,
\[
\gamma = \epsilon^2\Gamma,  \quad
\rho = \epsilon^2R,
\]
 we choose a fractional-power scaling law for $h$: 
\[
h = \epsilon^{3/2}H.
\]
This scaling will be shown to produce a balance of damping and driving 
terms at the leading order in the amplitude equation.
Expanding $\phi$ in powers of $\epsilon^{1/2}$,
\[
\phi = \phi_0  + \epsilon \phi_{1} + 
\epsilon^{3/2} \phi_{3/2} + \epsilon^2 \phi_2 + \epsilon^{5/2} \phi_{5/2} + \ldots,
\]
where $\phi_0 = \tanh X_0$,
and substituting in \eqref{12direq}, we obtain
Eq.\eqref{wobbmo} for 
$\phi_1$.
The partial differential equation arising at $\order{\epsilon^{3/2}}$ is
\[
\frac12 D_0^2\phi_{3/2} + \mathcal{L}\phi_{3/2} = 
\frac{H}{2}e^{i(\omega_0/2)T_0} + c.c.
\]
The solution $\phi_{3/2}$ to this equation has the form 
$\phi_{3/2}=\varphi_{3/2}^{(1/2)} e^{i (\omega_0/2) T_0}+c.c.$,
where the coefficient function $\varphi_{3/2}^{(1/2)}$ satisfies
the linear nonhomogeneous equation 
\begin{align*}
\left({\mathcal L}-\frac38 \right)\varphi_{3/2}^{(1/2)} &= \frac{H}{2}.
\end{align*}
Since $\frac38$ is not an eigenvalue of the
operator
${\mathcal L}$,  this equation has a unique bounded solution.
To determine it, we note that  
two homogeneous solutions of this equation, i.e. 
solutions of
$({\mathcal L}- 3/8)y=0$,
are given by Segur's formula 
\begin{multline}
\label{segurssolutions}
y_p(X_0) = \frac{1}{(1+ip)(2+ip)}e^{ipX_0}\\
\times (2-p^2-3ip \tanh X_0 - 3 \sech^2 X_0)
\end{multline}
 with $p= \pm i
\sqrt{13/4}$ \cite{Segur}.
Using these in the variation of parameters, we obtain
\begin{align*}
\varphi_{3/2}^{(1/2)} = \tfrac{4}{13}H(1-8\sech^2X_0).
\end{align*}
The term $\varphi_{3/2}^{(1/2)} e^{i(\omega_0/2)T_0}$ 
in the expansion of the wobbling kink represents the
background stationary wave induced by the driver.

The equations arising at $\order{\epsilon^2}$ are
the same as for the free wobbler and the $1:1$ parametric resonance,  
Eqs.\eqref{qua}-\eqref{F2a}. 
Hence the coefficients of the harmonic
components of $\phi_2$ are the same as in the undamped, 
undriven case. Namely, imposing the solvability conditions
$D_1V = 0$ and $D_1A = 0$,
we obtain \eqref{ph20} for
$\varphi_2^{(0)}$ 
and \eqref{ph22} for
$\varphi_2^{(2)}$. We also impose Eq.\eqref{par10}
to obtain $\varphi_2^{(1)}=0$.

At the order $\epsilon^{5/2}$ we have the equation
\[
\tfrac12 D_0^2 \phi_{5/2} + \mathcal{L} \phi_{5/2} = -6\phi_0\phi_1 \phi_{3/2} 
+ V D_0\partial_0 \phi_{3/2}.
\]
Its solution consists of the $\frac12$th and $\frac32$th harmonics
with the coefficient functions
\begin{align*}
\varphi_{5/2}^{(1/2)} &= HA u_a(X_0) + i\omega_0 HV u_b(X_0),\\
\varphi_{5/2}^{(3/2)} &= HA u_c(X_0),
\end{align*}
respectively. Here the functions $u_a$, $u_b$ and $u_c$ satisfy 
\begin{align*}
\left({\mathcal L}-\tfrac38 \right)u_a(X_0) &= -\tfrac{24}{13}(1-8\sech^2X_0)\sech X_0\tanh^2X_0,\\
\left({\mathcal L}-\tfrac38 \right)u_b(X_0) &=
\tfrac{32}{13}\sech^2X_0\tanh X_0,
\intertext{and}
\left({\mathcal L}-\tfrac{27}{8} \right)u_c(X_0) &= -\tfrac{24}{13}(1-8\sech^2X_0)\sech X_0\tanh^2X_0.
\end{align*}
 In order to determine $u_c(X_0)$ uniquely, we impose the radiation
boundary conditions. (These are necessary because
 the value $\frac{27}{8}$ lies in the continuous spectrum
of the operator $\cal L$.) The functions
$u_a$ and $u_c$ are even, while $u_b$ is odd. These three functions
can be easily found by solving the above nonhomogeneous
boundary-value problems numerically. 

Proceeding to the order $\epsilon^3$, we find
the equation \eqref{superharmeps3},
where $F_3$ is given by Eq.\eqref{F3p11} with the term $\frac12 H e^{i
\omega_0 T_0} \phi_0 + c.c.$ replaced with $-3\phi_0 \phi_{3/2}^2$. 
The solvability conditions for this equation are
\begin{equation}
D_2V = -2\Gamma V
\label{D2V}
\end{equation}
for the 0th harmonic, and
\begin{equation}
\label{superharmamp1}
D_2A = - \Gamma A - i\omega_0 R A + \tfrac{i}{2}  \zeta \omega_0 
 |A|^2A - \tfrac{i}{2} \omega_0   V^2A + \tfrac{60}{169}i\omega_0\pi H^2
\end{equation}
for the first harmonic. The latter equation includes both the 
damping and driving terms and so the resulting master equations 
could be expected to capture the essentials of the 
nearly-stationary wobbling of the kink
(i.e. wobbling in the vicinity of the fixed point of the amplitude
equations which arises due the balance of the damping and driving terms).
However the description provided by these amplitude
equations --- while being qualitatively correct ---
turns out to be insufficiently accurate  
 when  
  compared to numerical simulations of the full partial-differential
  equation \eqref{dd12}. (The source of this
  inaccuracy will be clarified below.) In search of greater accuracy, 
  we shall
proceed to higher orders.

The solution of Eq.\eqref{superharmeps3} has the form
\begin{multline*}
\phi_3=\varphi_3^{(0)} +
\varphi_3^{(1)} e^{i \omega_0 T_0} + c.c. \\
+ \varphi_3^{(2)}e^{2i \omega_0 T_0}+ c.c.
+ \varphi_3^{(3)}e^{3i \omega_0 T_0} + c.c.
\end{multline*}
The function $\varphi_3^{(2)}$ is
calculated
to be zero;
the coefficient of the 0th harmonic is given by
\begin{multline*}
\varphi_3^{(0)} = \tfrac{16}{169}H^2(45X_0\sech^2X_0
-3\tanh X_0\\-128\sech^2X_0\tanh X_0),
\end{multline*}
and the one for the first harmonic component is
\begin{multline}
\varphi_3^{(1)} = -\partial_2 A X_0\sech X_0 \tanh X_0
 + |A|^2A u_d (X_0) \\
 - \tfrac{2}{3}i\omega_0 (\Gamma+i\omega_0R)(3-4\sech^2X_0)
 + H^2 u_e (X_0),
\label{phi31super}
\end{multline}
where the functions $u_d(X_0)$ and $u_e(X_0)$ are the bounded solutions of
the following nonhomogeneous equations:
\begin{align}
({\mathcal L}-\tfrac32)u_d = &\tfrac{3}{2} \zeta \sech X_0\tanh X_0
+ 6\sech X_0\tanh^2 X_0
\nonumber
\\ \times\bigg[ 3X_0\sech^2&X_0
-\tfrac{5}{2}\sech^2X_0\tanh X_0 - f_1(X_0)\bigg],
\label{ud} \\
({\mathcal L}-\tfrac32 )u_e = &-\tfrac{48}{169}H^2\tanh
X_0(1-8\sech^2X_0)^2 \nonumber \\
&+\tfrac{180 }{169} \pi\sech X_0\tanh X_0. \nonumber
\end{align} 
 Since
$({\mathcal L}-3/2)$ is a parity-preserving operator 
while the right-hand sides of the above equations are given by odd functions,
and since the homogeneous solution $y_w= \sech X_0 \tanh X_0$ is also an odd function, 
it follows that the 
nonhomogeneous solutions $u_d$ and $u_e$ are both odd.
This is the only fact about $u_d$ and $u_e$ that we will
need in this section --- we do not need to know any detail of these functions here.
Nevertheless, we 
do evaluate the solution $u_d$ as it will be required later on, in the study
of the $1:1$ directly driven kink (section \ref{11direct}); we evaluate it 
using the variation of parameters and numerical 
integration. The nonhomogeneous solution is defined up to 
the addition of an arbitrary multiple of $y_w$; however this extra 
degree of freedom is fictitious as it can always be eliminated by a suitable
rescaling of $\epsilon$. [Accordingly, the extra term proportional to $y_w$
cancels in the integral $\eta$ where it appears in section \ref{11direct}
and does not contribute to the amplitude equations \eqref{directomega}].

To eliminate the quasisecular
term proportional to $X_0 \sech X_0 \tanh X_0$ in
\eqref{phi31super}, we set $\partial_2 A = 0$.

It will not be necessary to calculate the third harmonic component,
$\varphi_3^{(3)}$, 
as this does not contribute to the 0th or 1st harmonics at $\order{\epsilon^4}$, and
hence does not affect the $\epsilon^4$-correction to the 
amplitude equations.
Similarly, we shall not calculate $\phi_{7/2}$ as it only contains 
fractional harmonic components
which cannot impinge on the amplitude equations at $\order{\epsilon^4}$.
Hence we skip the order $\epsilon^{7/2}$.

At $\order{\epsilon^4}$, we obtain
\begin{equation}
\tfrac12 D_0^2 \phi_4+ {\cal L} \phi_4=F_4,
\label{eq_F4_12}
\end{equation}
where
\begin{widetext}
\begin{align*}
F_4 = (\partial_0\partial_1 - D_0 D_1)\phi_3 + (\partial_0\partial_2 - D_0 D_2)\phi_2 
+ \frac{1}{2}(\partial_1^2-D_1^2)\phi_2 + (\partial_0\partial_3 - D_0D_3)\phi_1 
+ (\partial_1 \partial_2-D_1D_2)\phi_1 \\ 
-3\phi_1^2\phi_2 
- 6\phi_0\phi_1\phi_3 
-3\phi_0\phi_2^2 + VD_0\partial_0\phi_3 +VD_0\partial_1\phi_2 
+ VD_0\partial_2\phi_1 +VD_1\partial_0\phi_2 +VD_1\partial_1\phi_1 +VD_2\partial_0\phi_1 \\
+ \frac{1}{2}D_2V\partial_0\phi_1 + \frac{1}{2}D_3V \partial_0\phi_0 
- \frac{1}{2}V^2\partial_0^2\phi_2 - V^2\partial_0\partial_1\phi_1 - \Gamma D_0\phi_2 
- \Gamma D_1\phi_1 + \Gamma V\partial_0\phi_1 - RD_0^2\phi_2 + VRD_0\partial_0\phi_1.
\end{align*}
\end{widetext}
The corresponding solvability conditions are
\begin{equation}
D_3V = 0
\label{D3V}
\end{equation}
and
\begin{align}
\label{superharmamp2}
D_3A = -\tfrac{1}{2}i\omega_0\lambda H^2 A
\end{align}
where
\begin{multline*}
\lambda = \int_{-\infty}^{\infty}\sech X_0 \tanh X_0 
\left[-\tfrac{96}{169}(45X_0\sech^2X_0 \right. \\
-3\tanh X_0-128\sech^2X_0\tanh X_0)\sech X_0\tanh^2X_0 \\
- \tfrac{24}{13}\tanh X_0(1-8\sech^2X_0)u_a(X_0) \\
- \tfrac{24}{13}\tanh X_0(1-8\sech^2X_0)u_c(X_0) \\
\left. - \tfrac{96}{169}\sech X_0\tanh X_0(1-8\sech^2X_0)^2 \right] dX_0. 
\end{multline*}
Numerically,
\[
\lambda= \lambda_R +i \lambda_I = -7.4656 - i 1.6785.
\]
 Expanding the derivative
$\partial/\partial \tau$ as $D_0 + \epsilon D_1 + \epsilon^2 D_2 +  \ldots$
and recalling that $d\tau/dt =  1+\rho$,
we combine  equations \eqref{superharmamp1} and 
\eqref{superharmamp2}. We also combine \eqref{D2V} and \eqref{D3V}.
This yields a system of two master equations:
\begin{subequations}
\label{directomegaov2}
\begin{align}
{\dot a} &= -\gamma a - i\omega_0 \rho a + i \frac{\omega_0 \zeta}{2}  |a|^2a
- i \frac{\omega_0}{2}  v^2a \nonumber \\
 &+ i\tfrac{60 }{169}\pi  \omega_0 h^2 -\tfrac{1}{2}i\omega_0\lambda h^2a 
+ \order{|a|^5}, \label{adir} \\
{\dot v} &= -2\gamma v + \order{|a|^5}. \label{vdir}
\end{align}
\end{subequations}
It is essential to combine the slow-scale equations in this way, 
rather than solving the individual equations 
with the assumption that the different scales are independent.  
Solving individual equations separately would be
 illegitimate because in integrating the equations one is covering 
more than one timescale. For example, solving
Eqs.\eqref{D3V} and \eqref{superharmamp2} we would be 
integrating over the scale $T_3$  which includes a shorter timescale 
$T_2$.

\begin{figure}
\includegraphics{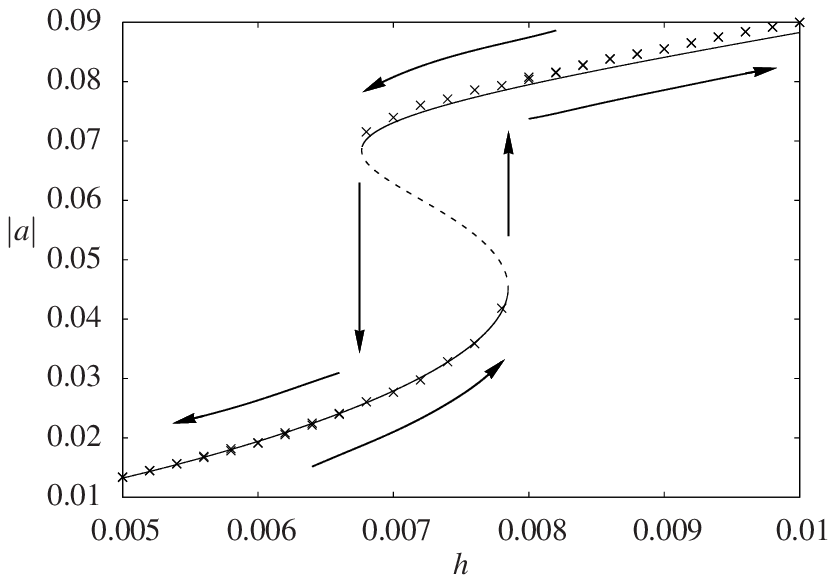}
\caption{\label{fig12dir} The hysteresis loop in the 1:2 directly driven
$\phi^4$ equation, Eq.\eqref{dd12}, with
$\gamma = 0.001$ and $\rho = -0.002$. The driving strength $h$
is increased from 0.005 to 0.01 in increments of 0.0002 
and then reduced back to 0.005 (as indicated by the arrows). Crosses mark
results of simulation of the PDE, Eq.\eqref{dd12}. The continuous and 
dashed lines depict the stable and unstable fixed points of the 
amplitude equation \eqref{adir} with $v=0$.}
\end{figure}

All terms in the right-hand side of Eq.\eqref{adir} are of the order
$|a|^3$, except the last term which is $\order{|a|^4}$. This last term 
is the correction coming from the 
fourth order of the perturbation expansion. As we have already
mentioned, the amplitude equations \eqref{directomegaov2} without 
this term produce an inaccurate description of the dynamics 
in the region of interest (i.e. in the vicinity of the fixed points). 
On the other hand, if the above fourth-order term is included,
the predictions of the amplitude equations \eqref{directomegaov2}
turn out to be in good agreement with results of the 
direct numerical simulations of the full partial differential equation 
\eqref{dd12} --- see Fig.\ref{fig12dir}. 
The substantial 
 improvement in accuracy
is due to the large value of the coefficient $\lambda$. 

\subsection{Reduced Two-Dimensional System}

Introducing the notation 
\[
\gamma'= \gamma- \frac{ \lambda_I}{2} \omega_0 h^2,
\quad
\rho'= \rho+ \frac{\lambda_R}{2} h^2, \quad h'=-\frac{480}{169}h^2,
\]
the amplitude equation \eqref{adir} can be written as 
 \begin{eqnarray}
{\dot a} = -\gamma' a - i\omega_0 \rho' a + i \frac{\zeta}{2} \omega_0 |a|^2a
-  \frac{i}{2} \omega_0 v^2a \nonumber \\
 - i \frac{\pi}{8}  \omega_0 h'  
+ \order{|a|^5},
\label{amodi}
\end{eqnarray}
 which has the same form as the amplitude equation for the $1:1$ parametric
resonance, Eq.\eqref{paramomegaamplitudea}.
Consequently, the dynamics of the $1:2$ directly driven wobbling kink 
will have some similarities
with the dynamics of the wobbler driven by the $1:1$ parametric
force.

According to Eq.\eqref{vdir},
the velocity will be damped until it is 
so small [$\order{|a|^3}$] that 
it can be disregarded in Eq.\eqref{adir}; 
hence after an initial transient the dynamics will be governed by 
Eq.\eqref{adir} with $v=0$.
Similarly to
Eq.\eqref{paramomegaamplitude} with $v=0$, equation \eqref{adir} does not
have closed orbits. All trajectories crossing the circle
$|a|= \frac{60 }{169}\pi \omega_0 h^2 / \gamma $ flow inwards
and so no trajectories can escape to infinity.
Therefore, all 
trajectories must flow towards fixed points. 
If $a$ is a fixed point, the absolute value $|a|$ satisfies
\begin{multline}
 h^4\left[ \frac14|\lambda|^2-\frac{\pi^2}{|a|^2}
 \left(\frac{60}{169}\right)^2\right] \\
+h^2\left[ \lambda_R \left(\rho-\frac12 \zeta_R|a|^2\right) - 
\lambda_I \left(\frac{\gamma}{\omega_0}+\frac12 \zeta_I|a|^2\right)\right] \\
+\left(\rho-\frac12 \zeta_R|a|^2\right)^2+\left(\frac{\gamma}{\omega_0}
+\frac12 \zeta_I|a|^2\right)^2 = 0.
\label{biqua}
\end{multline}
The left-hand side of this equation is
a bi-quadratic in $h$. Solving for $h$ we obtain 
an explicit expression for $h=h_{1,2}(|a|^2)$;
the roots $|a|^2$ are found by inverting these explicit functions
for each $\rho$, $\gamma$ and $h$. 
The lower branch of the  bi-quadratic 
is plotted in figure \ref{fig12dir} along with results
from the numerical simulations of 
the full PDE \eqref{dd12}. We note that, as in the
$1:1$ parametrically driven $\phi^4$ equation, there 
is no threshold driving strength for the
existence of the nonzero wobbling amplitude here.

Here it is appropriate to recall that $\rho$ and $\gamma$ 
were assumed to be of the same order while $h$ is $\order{\gamma^{3/4}}$. 
As in the case of the system \eqref{paramomegaamplitudea}, 
the absence or presence of hysteresis in 
the dynamics
depends on whether $\rho$ is above or below the critical value
$\rho_0(\gamma)$ given by Eq.\eqref{rhoga}. If the difference 
$\rho-\rho_0$ is positive and of order $\gamma$, there is only 
one root $|a|$ for each value of $h$.  The corresponding fixed point is
obviously stable. If, on the other hand, the difference 
$\rho-\rho_0$ is negative (but still of the order $\gamma$), we
have three
roots $|a|^2$ for each $h=\order{\gamma^{3/4}}$
in the interval $(h_+, h_-)$. These roots correspond 
to three fixed points, two of which are stable (see Fig.\ref{fig12dir}). 
The values $h_+$ and $h_-$ at which the subcritical bifurcations occur, are
given, approximately, by
\begin{equation}
h_\pm^2 = \frac{1}{{\frak c}_0} \left( P_\pm + \sqrt{P_\pm^2+ 2 {\frak
c}_0 Q_\pm}
\right),
\label{hpm_dir12}
\end{equation}
where
\begin{eqnarray*}
P_\pm =  \frac{1}{18} 
\left[  
{\frak c}_1(3 \beta^2- 5 \alpha^2)+ 6 \alpha \delta \right.
 \\ \left.  \mp 
(4{\frak c}_1 \alpha-3 \delta) \sqrt{\alpha^2- 3 \beta^2} \right],
 \\
Q_\pm =\frac{1}{27} \left[ \alpha(\alpha^2+ 9 \beta^2) \mp (\alpha^2- 3
\beta^2)^{3/2} \right],
 \\
{\frak c}_0=2 \left( \frac{30 \pi}{169} \right)^2 |\zeta|^4= 0.3280,
 \\
{\frak c}_1= \zeta_R \lambda_R + \zeta_I \lambda_I =6.2747,
 \\ 
\delta=|\zeta|^2 \left( \lambda_R \rho -\lambda_I \frac{\gamma}{\omega_0}
\right),
\end{eqnarray*}
and $\alpha$ and $\beta$ are defined by Eq.\eqref{alpha_beta}. 
For $\gamma=10^{-3}$ and $\rho=-2 \times 10^{-3}$, 
the bifurcation values obtained from 
Eq.\eqref{hpm_dir12} are $h_+ =6.74 \times 10^{-3}$
and $h_- =7.78 \times 10^{-3}$ while 
the numerical simulations of the full PDE give
 $6.6 \times 10^{-3}< h_+< 6.8 \times 10^{-3}$
and $7.8 \times 10^{-3}< h_-<8.0 \times 10^{-3}$.
For $\rho=-3 \times 10^{-3}$, the simulations 
show a more pronounced hysteresis loop whereas for
$\rho= -1 \times 10^{-3}$, the 
hysteresis was seen to disappear. (In both cases $\gamma$ was kept
at $10^{-3}$.) These observations 
are consistent with the value of $\rho_0$ given by the amplitude
equations.
[For $\gamma=10^{-3}$, Eq.\eqref{rhoga} gives $\rho_0=-1.1 \times
10^{-3}$.]

\subsection{1:2 Directly Driven Wobbler}

Finally, we 
produce the first several orders of the perturbation expansion
for the $1:2$ directly driven wobbling kink:
\begin{multline}
\phi(x,t) = \tanh [(1-3|a|^2)\xi]
 + a \sech \xi \tanh \xi  e^{i\Omega t} + c.c. \\
 + \tfrac{4}{13} h ( 1- 8 \sech^2 \xi) e^{i (\Omega/2) t} + c.c.\\ 
+ 2|a|^2 \sech^2\xi \tanh \xi + a^2 f_1(\xi )e^{2i \Omega t} + c.c. 
 + \order{|a|^{5/2}}. \label{12direxp}
\end{multline}
When $t$ is sufficiently large, the 
variable $\xi$ in this expression equals $x-x_0$ (where $x_0$ is a constant
determined by the initial conditions)
and $a$ is a stable fixed point of the
system \eqref{adir} with $v=0$ [a unique fixed point or one
of the two stable fixed points depending on whether $h$ is outside 
or inside the bistability interval $(h_+,h_-)$.]

The interpretation of terms in \eqref{12direxp} is the same as
in the previous sections. The frequency of the wobbling 
[where the wobbling mode
is given by
the sum of the third and second term in the first line in \eqref{12direxp}]
 is locked to double the 
driving frequency. The term proportional to $h$
in the second line describes a stationary wave induced by the 
driver. As in the previous sections, the expansion \eqref{12direxp}
is only valid at the length scale $|\xi|= \order{1}$. 
The standard analysis involving outer expansions demonstrates
that for larger distances, we have groups of second-harmonic
radiation waves moving away from the kink and leaving
in their wake  a sinusoidal waveform of constant amplitude.

\subsection{Qualitative Analysis}

The driving term 
$h \cos (\frac{\Omega}{2}t)$ is not in resonance with the natural frequency
of the wobbler,
nor does its parity coincide with the parity
of the wobbling mode. Therefore the ability of the 
$1:2$ direct driving to sustain the wobbling is surprising and
requires a qualitative explanation.

The authors of 
\cite{QuinteroLetter}
propose that the mechanism which brings about the 
unexpected superharmonic resonance is the coupling of the translation 
mode and the wobbling mode.
Our  explanation for this phenomenon is rather different 
and unrelated
to the translation mode.
The way the driver
 affects the wobbler is by exciting an 
even-parity standing wave ($\phi_{3/2}$) at the frequency $\Omega/2$ which then 
undergoes  nonlinear frequency doubling and parity 
transmutation
through the term $\epsilon^3 \phi_0 \phi_{3/2}^2$ in Eq.\eqref{dd12}. 
This latter term serves as an effective
driver to the wobbling mode; it has the resonant frequency
and ``correct" parity.  

 Since this mechanism involves 
a two-stage process and the resulting effective driving strength
is proportional to $h^2$, this type of driving produces a relatively
weak response.

\subsection{Chaotic Wobblers?}

The authors of Ref.\cite{QuinteroLetter,
QuinteroDirect} observed chaotic kink dynamics
in numerical simulations of the $1:2$ directly driven wobbling kink.
An indirect confirmation of the existence of
chaotic motions comes also from the collective-coordinate approach
 which predicts an unbounded growth of the kink's 
 width, energy and velocity at resonance
 \cite{QuinteroLetter,QuinteroDirect}. 
On the other hand, our amplitude equations \eqref{directomegaov2} 
with $\gamma \neq 0$ reduce to a two-dimensional
dynamical system which can obviously  not exhibit any chaotic attractors.

To find an explanation for this disagreement, we have 
carried out a series of numerical simulations of the partial
differential equation \eqref{dd12} at a range of 
driving strengths and damping coefficients.
In all our experiments, we confined ourselves to
zero detuning, $\rho=0$.
We could not detect any sign of chaotic dynamics
for  $h$ smaller than a certain minimum value, not even in the undamped case.
However for $h$ greater or equal than 0.05 and sufficiently
small $\gamma$, 
our numerical simulations 
did reveal kinks performing erratic motion,
where initially close profiles were seen to diverge exponentially fast.
For $h=0.05$, $0.06$ and $0.08$, chaos was observed 
in simulations with $\gamma$ smaller or equal to
$10^{-3}$, $2 \times 10^{-3}$, and $6 \times 10^{-3}$,
respectively,
whereas the same sequence  of $h$ values paired 
with $\gamma=2 \times 10^{-3}$,
$3 \times 10^{-3}$  and $7 \times 10^{-3}$,
respectively, did not feature any chaotic trajectories.
Therefore chaotic attractors may only arise when the 
damping is extremely weak, much weaker than $\order{h^{4/3}}$.
This is the reason why  the chaotic dynamics  is not captured by our 
amplitude equations 
\eqref{directomegaov2} which have
been derived on the assumption that
$h=\order{\epsilon^{3/2}}$ and $\gamma =\order{\epsilon^2}$.

The description of chaotic motions by means of  amplitude
equations is 
a topic of future research. We expect our asymptotic
method to remain applicable in this situation, with the
appropriate adjustment of the scaling laws of the $a$ and $v$
variables and parameters
of the damping and driving.

\section{$1:1$ direct resonance}
\label{11direct}

\subsection{Multiscale Expansion}

Finally, we explore the effect of the direct driving near the 
natural wobbling frequency of the kink. 
The equation
is 
\begin{equation}
\tfrac{1}{2}\phi_{tt}-\tfrac{1}{2}\phi_{xx} + \gamma \phi_t - \phi +
\phi^3 = h\cos(\Omega t),
\label{dd11}
\end{equation}
where $\Omega = \omega_0(1 + \rho)$. 
We let $a = \epsilon A$ and adopt the following scalings for the three small 
parameters:
\begin{equation}
h = \epsilon H,  \quad  \gamma = \epsilon^2\Gamma, \quad \rho = \epsilon^2R.
\label{sca_11d}
\end{equation}
This time, we assume that the velocity is scaled as
$v = \epsilon^2 V$ (and not as $v=\epsilon V$).  We
shall find a non-trivial evolution equation for $v$, and with the above
scalings, the leading-order dynamics of $a$ and $v$ will occur on the same
timescale.  While other scalings could be investigated, 
the variables $v$ and $a$ would
then change on different timescales and so would effectively be decoupled
for small $\epsilon$. Therefore, the chosen scalings 
correspond to the richest, three-dimensional,  dynamics.
Rescaling the time  so that $\Omega t = \omega_0 \tau$ and
transforming to the moving frame as in Eq.\eqref{xitau}, the equation
\eqref{dd11} becomes
\begin{multline}
\tfrac{1}{2} (1+\rho)^2 \phi_{\tau\tau} - v(1+\rho)\phi_{\xi\tau} 
- \frac{v_{\tau}}{2}(1+ \rho) \phi_{\xi} - \frac{1-v^2}{2}\phi_{\xi\xi} \\
- \phi + \phi^3 = h\cos(\omega_0 \tau )
+ v \gamma \phi_{\xi}-\gamma (1+\rho) \phi_{\tau}.
\label{d11A}
\end{multline}
We expand $\phi$ as in Eq.\eqref{phiexpans}.

With the driving amplitude of the order $\epsilon$, the linear
perturbation 
consists of the wobbling mode and a standing wave
excited by the driver:
\begin{align*}
\phi_1 &= \left[ A\sech X_0\tanh X_0+H(1-2\sech^2X_0)\right] e^{i\omega_0 T_0} + c.c.
\end{align*}
Next, at $\order{\epsilon^2}$, 
we obtain the equation \eqref{qua} with
\[
F_2=(\partial_0\partial_1 - D_0 D_1)\phi_1 
- 3\phi_0\phi_1^2.
\]
Once transients have died out, the solution to Eq.\eqref{qua} 
will consist only of 
the harmonics present in the forcing, i.e. will have the form 
\eqref{phi2m}. 
The solvability condition for 
the first-harmonic component
is $D_1A = 0$; assuming that this condition is in place, 
we obtain  $\varphi_2^{(1)}(X_0)= -\partial_1A X_0\sech X_0 \tanh X_0$. 
 To avoid the quasisecular behaviour at infinity
we impose  $\partial_1A = 0$,
which results in
\begin{equation}
\varphi_2^{(1)} = 0.
\label{11A1}
\end{equation}
 The other two harmonic components of
the quadratic correction  $\phi_2$ have the coefficients
\begin{multline}
\varphi_2^{(0)} = |A|^2\sech^2X_0(2 \tanh X_0 -3 X_0) \\
+ H^2(9X_0\sech^2X_0-3\tanh X_0-8\sech^2X_0\tanh X_0) \\
- 4H(A+\conj{A})\sech X_0(1+\sech^2X_0) 
\label{11A0}
\end{multline}
 and
\begin{equation}
\varphi_2^{(2)} = A^2 f_1(X_0) + A Hf_3(X_0) + H^2f_4(X_0),
\label{A11ex}
\end{equation}
where  $f_1$ is as in Eq.\eqref{f1},
and the functions $f_3$ and $f_4$ are defined by
\begin{multline}
f_3(X_0) = \tfrac{1}{2}\sech X_0 - 4\sech^3X_0 \\
- \tfrac{15}{32}i k_0(3-\tanh^2 X_0 +i k_0\tanh X_0)
[\conj{J}_1(X_0)-J_1^{\infty}] e^{i k_0 X_0} \\
  + \tfrac{15}{32}i k_0(3-\tanh^2 X_0 -ik_0\tanh X_0)J_1(X_0)e^{-i k_0 X_0}
  \label{f3}
\end{multline}
and
\be
f_4(X_0)=-\frac72 f_1(X_0) + \frac14 \tanh X_0 (3-2 \sech^2 X_0).
\label{f4}
\ee
The function $J_1(X_0)$ is given by the integral \eqref{Jdef}
with $n=1$.
One can show that $f_3(X_0)$ is an even function and $f_4(X_0)$ is odd.

We note a quasisecular term $(9H^2-3|A|^2) X_0 \sech^2 X_0$ 
in Eq.\eqref{11A0}; this term does not lead to the nonuniformity of the
expansion as it can be incorporated in the variable width of the
kink.
[See Eq.\eqref{dd11ex} below.]

The partial differential
equation arising at the order $\epsilon^3$, is Eq.\eqref{superharmeps3},
with $F_3$ given by 
\begin{widetext}
\begin{eqnarray*}
F_3 = (\partial_0 \partial_1 - D_0 D_1) \phi_2 
+ (\partial_0 \partial_2 - D_0 D_2) \phi_1
+ \tfrac{1}{2}(\partial_1^2 - D_1^2)\phi_1 
- \phi_1^3  \\ - 6 \phi_0 \phi_1 \phi_2 
+ V D_0 \partial_0 \phi_1 + \tfrac{1}{2} D_1 V \partial_0 \phi_0
- \Gamma D_0 \phi_1 - RD_0^2 \phi_1. 
\end{eqnarray*}
\end{widetext}
The solvability conditions give rise to 
amplitude equations: 
\begin{equation}
D_1V = 0 \label{nodamp}
\end{equation}
for the zeroth harmonic, and 
\begin{multline}
\label{firstampeqomega}
D_2A + \Gamma A +i\omega_0 RA - \tfrac{1}{2} i\omega_0 \zeta |A|^2A
- \tfrac{3}{4}\pi VH \\
 + \tfrac{1}{2} i\omega_0 \nu H^2A 
+ \tfrac{1}{2} i\omega_0 \mu H^2\conj{A} = 0
\end{multline}
for the first harmonic. In Eq.\eqref{firstampeqomega},
we have introduced
\begin{widetext}
\begin{eqnarray*}
\nu = \int_{-\infty}^{\infty} \sech X_0 \tanh X_0 \left[ -6\tanh X_0 
(1-2\sech^2X_0)  
 f_3(X_0) - 6\sech X_0\tanh^2 X_0 (9X_0\sech^2X_0 - 
3\tanh X_0 \right. \\ \left. - 8\sech^2X_0\tanh X_0)  + 24\tanh X_0(\sech X_0 + \sech^3X_0)
(1 - 2\sech^2X_0)  - 6\sech X_0\tanh X_0(1-2\sech^2X_0)^2   \right] 
dX_0,
\end{eqnarray*}
and 
\begin{eqnarray*}
\mu = \int_{-\infty}^{\infty} \sech X_0\tanh X_0 
\left[ 24\tanh X_0(1-2\sech^2X_0)(\sech X_0+\sech^3X_0) \right. \\ 
- 3\sech X_0\tanh X_0(1-2\sech^2X_0)^2   \left. - 6\sech X_0\tanh^2X_0 f_4(X_0)
 \right] dX_0.
\end{eqnarray*}
\end{widetext}
Numerically,
\begin{equation*}
\nu= 4.159 - i 0.3258, \quad \mu =1.022 + i 0.1623.
\end{equation*}

We note that the velocity enters 
the amplitude equation \eqref{firstampeqomega} as a coefficient
in front of one of its two driving terms. 
On the other hand, Eq.\eqref{nodamp} 
implies that $V$ does not tend to zero --- at
least on the timescale $T_1$. 
 In order to check whether the velocity decays on a longer
 timescale and hence whether the translational motion can 
 drive the wobbling,
we take the expansion to higher orders.  

The cubic correction has the form
\begin{multline*}
\phi_3=\varphi_3^{(0)} +
\varphi_3^{(1)} e^{i \omega_0 T_0} + c.c. \\
+ \varphi_3^{(2)}e^{2i \omega_0 T_0}+ c.c.
+ \varphi_3^{(3)}e^{3i \omega_0 T_0} + c.c.
\end{multline*}
The 0th harmonic component,
$\varphi_3^{(0)}$, is evaluated to be zero,
and the coefficient of the first-harmonic component is
\begin{multline}
\varphi_3^{(1)} =|A|^2A u_d(X_0)  \\ 
-(\partial_2 A + i\omega_0 V A)X_0\sech X_0 \tanh X_0 
  \\ + i\omega_0 VH u_1(X_0)
 + H|A|^2 u_2(X_0) + HA^2u_3(X_0)  \\ + H A^2 u_4(X_0)
  - \tfrac{2}{3}i\omega_0 (\Gamma+i\omega_0R)(3-4\sech^2X_0) \\
 + H^2A u_5(X_0) + H^2\conj{A} u_6(X_0) + H^3 u_7(X_0).
\label{phi31}
\end{multline} 
Here $u_d(X_0)$ was defined in the 
previous section as the bounded solution of Eq.\eqref{ud}, and 
the functions $u_n(X_0)$ ($n =1,...,7$)  are the bounded solutions of
the following nonhomogeneous equations:
\begin{widetext}
\begin{align*}
({\mathcal L}-3/2)u_1 &= -4\sech^2X_0\tanh X_0
+ \frac{3\pi}{4}\sech X_0\tanh X_0, \\
({\mathcal L}-3/2)u_2 &= -6\sech^2X_0\tanh^2X_0(1-2\sech^2X_0)
+24\sech^2X_0\tanh^2X_0(1+\sech^2X_0)\\
&\;\;-6\sech^2X_0\tanh X_0(2\tanh X_0-3X_0)(1-2\sech^2X_0)
-6\sech X_0\tanh^2 X_0 f_3(X_0),\\
({\mathcal L}-3/2)u_3 &= 3\sech^2X_0\tanh^2X_0(7+10\sech^2X_0)
+12\sech^2X_0\tanh X_0f_1(X_0),\\
({\mathcal L}-3/2)u_4 &= -6\tanh X_0f_1(X_0), \\
({\mathcal L}-3/2)u_5 &= -6\tanh X_0(1-2\sech^2X_0)f_3(X_0)
+24\sech X_0\tanh X_0(1+\sech^2X_0)(1-2\sech^2X_0) \\
&\;\;-6\sech X_0\tanh^2X_0(9X_0\sech^2X_0-3\tanh X_0-8\sech^2X_0\tanh X_0) \\
&\;\;-6\sech X_0\tanh X_0(1-2\sech^2X_0)^2 - \tfrac{3}{2}\nu\sech X_0\tanh X_0, \\
({\mathcal L}-3/2)u_6 &= -6\sech X_0\tanh^2X_0f_4(X_0)
+24\sech X_0\tanh X_0(1+\sech^2X_0)(1-2\sech^2X_0) \\
&\;\;-3\sech X_0\tanh X_0(1-2\sech^2X_0)^2 - \tfrac{3}{2}\mu\sech X_0\tanh X_0, \\
\intertext{and}
({\mathcal L}-3/2)u_7 &= -6\tanh X_0(1-2\sech^2X_0)f_4(X_0) -3(1-2\sech^2X_0)^3 \\
&\;\;-6\sech X_0\tanh^2X_0(9X_0\sech^2X_0-3\tanh X_0-8\sech^2X_0\tanh X_0).
\end{align*}
\end{widetext}
Like the functions $u_d$ and $u_e$ of the previous section, 
the solutions $u_n(X_0)$ are defined up to the addition of a multiple of $y_w$.
As in the previous section, this does not provide any
extra degrees of freedom and the multiple of $y_w$ cancels out in the 
integrals $\eta$ and $\chi$ below. 
The solutions
$u_1$, $u_5$, and $u_6$ are odd, while $u_2$, $u_3$, $u_4$ and $u_7$
can be chosen to be even functions.
The only fact about $u_1$ that we need is that it is a real solution;
owing to its reality, $u_1$ does not contribute to the solvability conditions 
below. The solutions $u_n(X_0)$ with $n=2,..., 7$ are determined 
using the  variation of parameters and numerical 
integration.

To eliminate the quasisecular
term proportional to $X_0 \sech X_0 \tanh X_0$ in
\eqref{phi31}, we set $\partial_2 A = -i\omega_0 VA$.

It is not necessary to calculate the 
second and third harmonic components, $\varphi_3^{(2)}$ and
$\varphi_3^{(3)}$,
as these do not contribute to the zeroth harmonic 
at fourth order in $\epsilon$,
where the leading order behaviour of $V$ will reveal itself.
The equation arising at ${\cal O}(\epsilon^4)$ is Eq.\eqref{eq_F4_12}, with
\begin{widetext}
\begin{align*}
 F_4 = (\partial_0\partial_1 - D_0 D_1)\phi_3 + (\partial_0\partial_2 - D_0 D_2)\phi_2 
+ \frac{1}{2}(\partial_1^2-D_1^2)\phi_2 + (\partial_0\partial_3 - D_0D_3)\phi_1
 + (\partial_1\partial_2-D_1D_2)\phi_1  \\
-3\phi_1^2\phi_2 - 6\phi_0\phi_1\phi_3 -3\phi_0\phi_2^2 + VD_0\partial_0\phi_2 +VD_0\partial_1\phi_1
 + \frac{1}{2}D_2V\partial_0\phi_0 - \frac{1}{2}V^2\partial_0^2\phi_0
- RD_0^2\phi_2 - \Gamma D_0\phi_2 + \Gamma V \partial_0\phi_0.
\end{align*}
The solvability condition for the zeroth harmonic
yields 
\begin{equation}
\label{D2Vdir} 
D_2V = -2\Gamma V - \tfrac{3}{2}\eta H |A|^2A + c.c.
- \tfrac{3}{2}\chi H^3A + c.c. \\
+ \tfrac{3\pi}{4}i\omega_0 H ( \Gamma  - i\omega_0 R)A +
c.c.,
\end{equation}
where
\begin{eqnarray*}
\eta = \int_{-\infty}^{\infty} \sech^2X_0 \left[ 24\sech^2X_0
\tanh^2X_0(\sech X_0+\sech^3X_0) - 3\sech^2X_0\tanh^2X_0\conj{f_3}(X_0)
\right.\\ 
- 6\sech^3X_0 \tanh X_0  (2\tanh X_0-3X_0)(1-2\sech^2X_0) - 6\sech X_0\tanh X_0
(1-2\sech^2X_0)f_1(X_0) \\ - 6\sech X_0\tanh^2X_0 u_3(X_0)  
 - 6\sech X_0\tanh^2X_0 u_4(X_0) \\
 -6\sech X_0\tanh^2X_0\conj{u_2}(X_0) + 24\sech^2X_0\tanh X_0(2\tanh
 X_0-3X_0) \\ \left. \times
 (\sech X_0+\sech^3X_0) 
- 6\tanh X_0(1-2\sech^2X_0)u_d(X_0) - 6\tanh X_0f_1(X_0)\conj{f_3}(X_0)
\right] dX_0,
\end{eqnarray*}
and
\begin{eqnarray*}
\chi = \int_{-\infty}^{\infty} \sech^2X_0 \left[
-6\sech X_0\tanh^2X_0\conj{u_7}(X_0) - 6\tanh
X_0(1-2\sech^2X_0)u_5(X_0) \right. \\
- 6\tanh X_0(1-2\sech^2X_0)\conj{u_6}(X_0) +
24(1-2\sech^2X_0)^2(\sech X_0+\sech^3X_0) \\ 
+ 18 \sech X_0 \tanh X_0(1+2\sech^2 X_0) \times
(9X_0\sech^2X_0-3\tanh X_0-8\sech^2X_0\tanh X_0)
\\ \left. 
 - 3(1-2\sech^2X_0)^2f_3(X_0) - 6\tanh X_0f_3(X_0)\conj{f_4}(X_0) 
 - 6\sech X_0\tanh X_0(1-2\sech^2X_0)\conj{f_4}(X_0) \right] dX_0.
\end{eqnarray*}
\end{widetext}
Numerically, 
\[
\eta = -2.005 - i0.3823, \quad
 \chi = -12.21 - i0.5706. 
\]

Writing $a = \epsilon A$ as before, 
and combining $D_1A=0$ with Eq.\eqref{firstampeqomega} with the help
of the chain rule, we obtain the amplitude
equation in terms of the unscaled parameters:
\begin{subequations}
\label{directomega}
\begin{multline}
\label{directomegaa}
{\dot a} = - \gamma a - i\omega_0 \rho a + \tfrac{1}{2} i\omega_0 \zeta |a|^2a
+ \tfrac{3\pi}{4} vh \\
 - \tfrac{1}{2} i\omega_0 \nu h^2a 
- \tfrac{1}{2} i\omega_0 \mu h^2\conj{a} + \order{|a|^5}. \\
\end{multline}
Similarly, combining Eq.\eqref{nodamp} with \eqref{D2Vdir}, 
we arrive at 
\begin{multline}
\label{directomegav}
{\dot v} = -2\gamma v - \tfrac{3}{2}\eta h |a|^2a + c.c.
- \tfrac{3}{2}\chi h^3a + c.c. \\
+ \tfrac{3\pi}{4}i\omega_0 h \left( \gamma  - i\omega_0 \rho \right)a  + c.c.
+\order{|a|^5}.
\end{multline}
\end{subequations}

\subsection{Reduced Dynamics in Three Dimensions}

\begin{figure}
\includegraphics{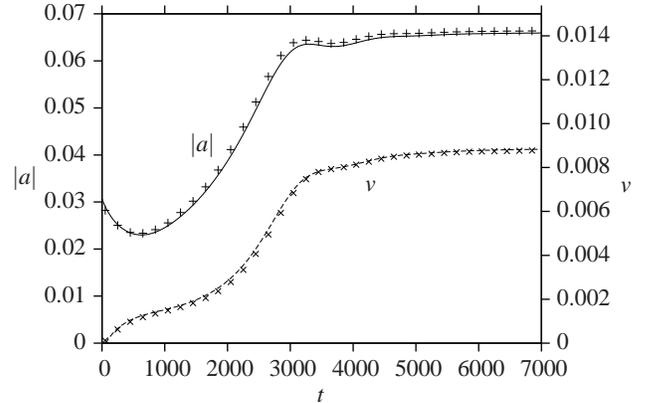}
\caption{\label{tseries1} An example of the kink being accelerated by
the $1:1$ direct driving. Here $h = 0.012$, $\gamma = 
1\times 10^{-3}$ and $\rho = 0$.  The crosses are measured values 
from numerical simulations of the original PDE, while the lines are 
the predictions of the amplitude
equations.  }
\end{figure}

Thanks to the $a$-dependent driving terms in Eq.\eqref{directomegav},
the direct driving can sustain the translational motion of the kink --- in 
contrast to the parametrically driven cases we have considered. 
Fig.\ref{tseries1} shows an example of
the kink accelerated by the 1:1 direct driving force which
simultaneously excites the wobbling. Note that results from
the three-dimensional system \eqref{directomega} are in excellent
agreement with predictions of the full partial differential equation
--- not only after the dynamics have settled to a stationary regime
but also during the transient phase.

\begin{figure}
\includegraphics{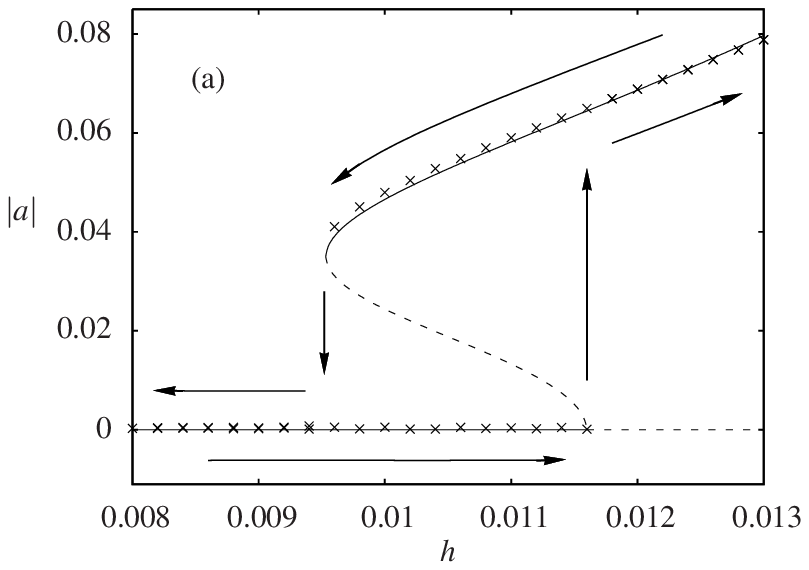}\qquad \includegraphics{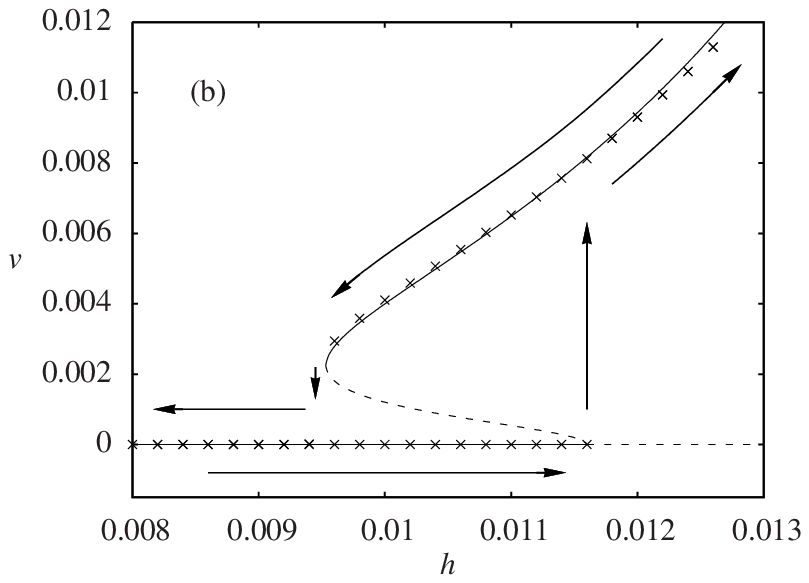}
\caption{\label{gammaplot}
The hysteresis loop in the
$1:1$ directly driven $\phi^4$ equation
with $\gamma =  10^{-3}$ and $\rho=-10^{-4}$.
The driving strength is increased from 
$h=8 \times 10^{-3}$ to $13 \times 10^{-3}$ in
increments of $0.2 \times 10^{-3}$ and then reduced back to $8 \times 10^{-3}$.
The ``crosses"
are measured values from numerical simulations of the  
partial differential equation \eqref{dd11}
whereas the continuous and dashed lines show the 
stable and unstable fixed points of the amplitude equations
\eqref{directomega}.
}
\end{figure}

The detailed analysis of the three-dimensional
system  \eqref{directomega} will be
reported elsewhere; here, we limit ourselves
to several basic observations. 
Firstly, it is straightforward to see that
when $h=0$, the trivial fixed point $a=v=0$ is the only 
attractor available in the system. 
Secondly, numerical simulations 
show that as $h$ is increased
for  fixed $\gamma$ and $\rho$, a nontrivial fixed point
bifurcates from the point $a=v=0$
[see Fig.\ref{gammaplot} (a,b)].
For lower values
of $\rho$, the bifurcation is subcritical 
 (as in the case shown in Fig.\ref{gammaplot}); for higher $\rho$, it
 is
supercritical. As $h$ approaches some critical driving
strength $h_{\rm c}$, both the
$|a|$- and $v$-component of the nontrivial fixed point tend to infinity. 
Finally, in the region $h>h_{\rm c}$, there are no stable fixed
points. In this region,
simulations of the system \eqref{directomega}
reveal a blow-up regime, where the functions
$|a(t)|, v(t)$ grow without bounds. 

To determine the critical value $h_{\rm c}$, we assume that 
the blow-up regime is self-similar, that is, that 
the growth of $v$ is pegged to that of $a$. This assumption
can 
be formalised by expanding $v$  and $\mbox{Arg} (a)$  in powers of
large  $|a|$:
\begin{subequations}
\label{self}
\begin{eqnarray}
v=V_3|a|^3 + V_1 |a|+ V_{-1}|a|^{-1}+...,
\label{selfa} \\
a=|a|e^{-i \theta}, \quad
\theta=\theta_0+\theta_{-2} |a|^{-2}+\theta_{-4} |a|^{-4}+ ... .
\label{selfb}
\end{eqnarray}
\end{subequations}  
Substituting these expansions  in Eq.\eqref{directomegaa}
and \eqref{directomegav},
and equating coefficients of like powers of $|a|$ to zero,
we can evaluate the coefficients $V_n$ and
$\theta_n$ to any order --- this justifies the assumption.
 
In particular, setting to zero the coefficients of
$|a|^3$ in Eq.\eqref{directomegaa}, gives
\begin{equation} 
e^{i \theta_0}= -i \frac{\zeta}{|\zeta|}, 
\quad
V_3=\frac{2\omega_0 |\zeta|}{3 \pi} \frac{1}{h}.
\label{V3}
\end{equation}
On the other hand, substituting Eqs.\eqref{self} in 
Eq.\eqref{directomegav}, we get an equation 
describing the growth of $|a(t)|$:
\begin{equation*}
\frac{d}{dt} |a|= {\frak r} |a| +{\cal O}(|a|^{-1}),
\end{equation*}
where the growth rate
\[
{\frak r}=-\frac23 \gamma -
\frac{\eta e^{-i\theta_0}+ \eta^* e^{i\theta_0}}{2} \frac{h}{V_3}.
\]
Substituting for $e^{i\theta_0}$ and $V_3$ from \eqref{V3},
this becomes
\begin{equation}
{\frak r}= -\frac23 \gamma -\frac{3 \pi}{2 \omega_0 |\zeta|} 
\frac{\eta e^{-i\theta_0}+ \eta^* e^{i\theta_0}}{2} 
h^2.
\label{brack}
\end{equation}
The growth of $|a|$ is due to the $h^2$ term in Eq.\eqref{brack}
which has a positive coefficient; the growth is damped by the 
$\gamma$ term. 
If we reduce $h$ keeping $\gamma$ fixed, then, at  $h=h_{\rm c}$
where
\begin{equation}
h_{\rm c}= \left[ \frac{8 \omega_0}{9 \pi}
\frac{|\zeta|^2}{i(\zeta \eta^* - \zeta^* \eta)} 
\right]^{1/2}
\gamma^{1/2}=0.6523 \gamma^{1/2},
\label{threshold}
\end{equation}
the growth rate will become equal to zero. At this point
the blow-up regime is replaced by a stable fixed point
--- which, however, still has large values of $|a|$ and $v$.
[Reducing $h$ further, the fixed point will persist but
the similarity relations \eqref{self} will no longer be valid.]  
Note that the critical value \eqref{threshold} does not depend on $\rho$.
Numerical simulations of  Eqs.\eqref{directomega}
carried out for a variety of $\gamma$ and $\rho$, 
reproduce the value of $h_{\rm c}$ to high accuracy.

\begin{figure}
\includegraphics{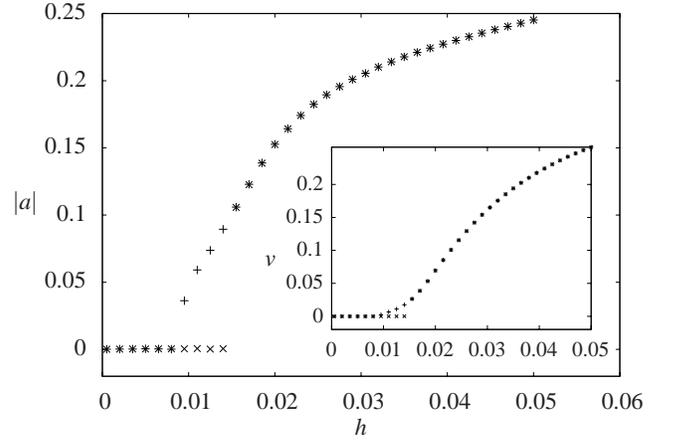}
\caption{\label{beyond}  Results of 
numerical simulations of Eq.\eqref{dd11} with $h$ raised beyond the
critical
value $h_{\rm c}=0.021$.
As in Fig.\ref{gammaplot}, in this plot $\gamma=10^{-3}$ and $\rho=-10^{-4}$.
The main panel shows values of $|a|$ and the 
inset values of $v$ as a function of $h$. 
The ``crosses" represent measurements obtained as
$h$ is increased from $0.0005$ to
$0.05$ in steps 
of $1.5 \times 10^{-3}$; the ``pluses" are obtained as $h$ is decreased
back to $0.0005$. An ``asterisk" results 
when a ``plus" is superimposed over the ``cross" at the 
same point.}
\end{figure}

As $h$ is increased towards $h_{\rm c}$ and neither
$|a|$- nor  $v$-components
of the stationary point are small any longer, the system 
\eqref{directomega} ceases to provide any reliable description for 
the dynamics of the wobbler. A natural question that arises here,
is what dynamical regime the kink settles to for $h$ just below $h_{\rm c}$
and for $h$ above $h_{\rm c}$. In other words, we want to know
what happens when the  the wobbler is driven with the small strength 
 $h$ of order $\gamma^{1/2}$ [so that the 
 conditions \eqref{sca_11d} are still in place] 
 for which our finite-dimensional approximation is no longer valid.
  To answer this,
we have conducted a series of numerical simulations of 
 Eq.\eqref{dd11} with $h$ raised from $h<h_{\rm c}$ to
 $h> h_{\rm c}$. The simulations reveal that in the region 
 inaccessible to 
our finite-dimensional approximation, the kink settles to wobbling with
a constant amplitude which is accompanied by its translational motion
with a constant velocity.
[See Fig.\ref{beyond}.] The numerically detected values of 
$a$ and $v$ are of order $\gamma^{1/3}$ in this region; this 
accounts for the inadequacy
 of our approximation which 
was based on the assumptions  $a=\order{\gamma^{1/2}}$
and $v=\order{\gamma}$. 
  
The behaviour of the $1:1$ externally  driven wobbling kink
above (and just below) the critical value 
is an issue to which we are planning to return
in the near future. 
To derive the correct set of amplitude equations, we
 will need to use our asymptotic approach with modified scalings
 for $a$ and $v$.

\subsection{1:1 Directly Driven Wobbler and Its Radiation}

Up to $\order{\epsilon^2}$, the perturbation expansion gives
\begin{multline}
\phi(x,t) = \tanh \left[ (1-3|a|^2+9h^2) \xi \right]
+ \big[a \sech \xi \tanh \xi \\ + h(1-2\sech^2\xi)\big] e^{i\Omega t} + c.c. 
+ 2|a|^2\sech^2 \xi \tanh \xi \\
- 4h(a+\conj{a}) \sech \xi \, (1+\sech^2\xi) \\
- h^2(3\tanh \xi+8\sech^2\xi\tanh \xi) \\
+ \left[a^2f_1(\xi) + haf_3(\xi) + h^2f_4(\xi)\right] e^{2i\Omega t} 
+ c.c. + \order{|a|^3}.
\label{dd11ex}
\end{multline}
For sufficiently large $t$, the variable $\xi$ is given by $x-vt-x_0$,
where $x_0$ is determined by the initial conditions.
The complex constant $a$
and the real $v$ are components of a stable fixed point (trivial or nontrivial)
of the system \eqref{directomega}. The functions $f_1$, $f_3$ and $f_4$
are given by Eqs.\eqref{f1}, \eqref{f3} and \eqref{f4}.
As in all previously considered driving
regimes, the $1:1$ direct driver excites a standing wave 
with the amplitude proportional to
the driver's strength and frequency equal to the frequency of
the driving [first two terms in the second line in \eqref{dd11ex}].
The standing wave includes also the second and zeroth harmonic,
both with the amplitudes of order $h^2$ (terms in the fourth and the
last line). 

Like the expansions in the previous
sections, Eq.\eqref{dd11ex} is only valid
at distances $|\xi|=\order{1}$. To describe the waveform at longer
distances, we consider the outer expansions
\[
\phi= \pm 1+ \epsilon (H e^{i \omega_0 T_0} + c.c.)
 + \epsilon^2 \phi_2 + \epsilon^3 \phi_3+
\dots
\]
in the regions $X_0>0$ and $X_0<0$, respectively. 
Substituting in the equation \eqref{d11A}, the order $\epsilon^2$ gives
\begin{eqnarray}
\phi_2= \mp 3H^2 \pm \frac34 H^2 e^{2 i\omega_0 T_0} + c.c.
\nonumber \\
+{\cal J}_\pm B_\pm e^{i(\omega_\pm T_0-k_\pm X_0)}  + c.c. \ .
\label{11A2}
\end{eqnarray}
Here the top and bottom sign pertain to the 
$X_0>0$ and $X_0<0$ region, respectively.
The amplitudes $B_\pm$ are functions of the
``slow" variables: $B_\pm= B_\pm(X_1,...;T_1,...)$
 and the
normalisation coefficients
${\cal J}_\pm$  have been introduced for later convenience.
Matching the outer solution \eqref{11A2} to the inner solution \eqref{phi2m} 
with coefficients as in 
\eqref{11A1}, \eqref{11A0} and \eqref{A11ex}, 
and choosing ${\cal J}_\pm =
\pm (2-ik_0)$,
we 
obtain $\omega_\pm= 2 \omega_0$, $k_\pm=\pm k_0$ and
\begin{widetext}
\begin{equation}
\label{BBA}
B_\pm (0,0,...;T_1,T_2,...)= \frac18 J_2^\infty \left[ 
A^2(0,0,...;T_2,T_3,...)
-\frac{7}{2} H^2 \right]  \pm i \frac{15}{32} 
k_0 J_1^\infty H A(0,0,...;T_2,T_3,...).
\end{equation}
\end{widetext}
Equations \eqref{BBA} are the boundary conditions for the 
amplitude fields $B_+$ and $B_-$. The equations of motion 
for these fields are obtained at the order $\epsilon^3$ of the 
outer expansion. Namely, the solvability conditions for $\varphi_3^{(2)}$, the 
coefficient of the second harmonic at the order $\epsilon^3$, give
\begin{subequations} \label{trans11}
\begin{eqnarray}
D_1 B_+ + c_0 \partial_1 B_+=0, \quad X_1>0  \\
D_1 B_- - c_0 \partial_1 B_-=0, \quad X_1<0,
\end{eqnarray}
\end{subequations}
where $c_0= k_0/(2 \omega_0)$.

As before, the analysis of the linear transport equations 
\eqref{trans11} under the boundary conditions
\eqref{BBA} is straightforward. The initial condition
$B_+(X_1,...;0,...)$ defined in $X_1>0$, propagates, unchanged, 
to the right and the initial condition
$B_-(X_1,...;0,...)$ defined in $X_1<0$, propagates, unchanged, 
to the left, both with the velocity $c_0$.
In the expanding region $-c_0 T_1<X_1<c_0 T_1$, the amplitudes $B_\pm$ are 
constants defined by the conditions \eqref{BBA}.  
In terms of the second-harmonic radiation, this
corresponds to two 
groups of radiation waves, diverging to the left and to the right,
and leaving a sinusoidal waveform of constant amplitude
in between. 
 
\subsection{Qualitative Analysis}
\label{Discussion}

The reason why the external force does not couple directly 
to the wobbling mode in the case of the $1:1$ external driving 
(as it did in the case of the $1:1$ {\it parametric\/} resonance), 
is the discrepancy in the parity of the driving 
profile and the wobbling mode. Instead, there are three
indirect amplification mechanisms at work in this case.  
In each of these, the central role is played by the even-parity
standing wave excited by the driver. Firstly, the square of
this standing wave couples to the wobbling mode via the term
$\epsilon^3 \phi_1^3$. Secondly, the second and zeroth harmonic of the standing
wave as 
well as the second-harmonic radiation excited by the standing wave
couple to the wobbling mode via the term $\epsilon^3 \phi_0 \phi_1 \phi_2$.
Thirdly, when the kink moves relative to the 
standing wave, the odd-parity wobbling mode acquires an
even parity component which then couples to the standing wave.
[This process is accounted for by the term $\epsilon^3 V D_0 \partial_0 \phi_1$
in Eq.\eqref{d11A}.]
Since the first two mechanisms rely upon a quadratic superharmonic 
of the induced standing wave
(with the amplitude of the superharmonic 
being proportional to $h^2$), 
and since the velocity of the kink (which determines the 
amplitude of the even component of the wobbling mode in
the third mechanism)
 is
small, the $1:1$ direct resonance is weak.

\section{Concluding remarks}
\label{Conclusions}

 In this paper, we have used the asymptotic method to study 
the wobbling kink driven by four types of resonant 
driving force, {\it viz.}, the $1:1$ and $2:1$ parametric, and $1:2$ and $1:1$ 
external driving. We have demonstrated 
the existence of resonance (i.e. the existence of 
sustained wobbling 
with nondecaying amplitude despite losing energy to radiation 
and dissipation) in all four cases. This conclusion (verified in direct
numerical simulations of the corresponding
partial differential equation) agrees with results of  
Quintero, S\'anchez and Mertens 
 who also demonstrated the existence of the resonance in the 
 $1:1$ parametrically and $1:2$ directly driven $\phi^4$ 
 equation
  \cite{QuinteroLetter,QuinteroDirect,QuinteroParametric}.
  However, we are in disagreement with these authors on the $1:1$ directly
  driven, damped equation. Namely, our method does capture the resonance 
  in this case whereas their collective coordinate approach does not. 
  (In fairness to 
the pioneering work of Quintero, S\'anchez and Mertens,  
their direct numerical simulations did reveal a resonant peak 
at the frequency of the driver
equal to the natural wobbling frequency of the kink  --- an experimental
result which, however, did not reconcile with their collective coordinate
predictions \cite{QuinteroLetter,QuinteroDirect}.)

In each of the four driving regimes that we have considered in this
paper, we have derived a system of equations for the 
complex amplitude of the wobbling coupled to the velocity
of the kink. The predictions based on this dynamical
system are in agreement with results of the
direct numerical simulation 
of the full partial differential equation. 
In three out of four cases considered,
the velocity of the kink is shown to decay to zero as time
advances, as a result of which the dimension of 
this dynamical system reduces from 3 to 2. Only in one case
(the case of the $1:1$ directly driven wobbler) 
does the velocity of the kink not necessarily decay to zero.
In this
latter case the wobbling of the kink is accompanied by 
its motion with nonzero velocity.

Each of the four dynamical systems derived here
give rise to a bifurcation diagram featuring
 bistability and hysteretical transitions in the
 wobbling amplitude. 
 In the $1:1$ parametric and $1:2$ direct resonances, 
 the bistability is between two nonzero values of the 
 wobbling amplitude, whereas in the $2:1$ parametrically
 and $1:1$ directly driven $\phi^4$ equations, one of
 the two stable regimes involves a nonzero and the other one a
 zero amplitude. It is fitting to note here that the 
 collective coordinate approach  
 \cite{QuinteroLetter,QuinteroDirect,QuinteroParametric} 
 does not capture
 the bistability and hysteresis.

In section \ref{h_vs_sh}, we ranked 
the two parametric resonances according to the 
amplitude of the stationary wobbling resulting from the
driving with a certain reference strength, $h$. 
Adding to this hierarchy the two
direct resonances produces the following ranking.
The  $1:1$ parametric resonance is the strongest
of the four cases; in this
case the amplitude of the stationary wobbling, $a$,
is of the order $h^{1/3}$. The $2:1$ parametric resonance is 
second strongest; in this case 
the kink
responds with the wobbling amplitude $a \sim h^{1/2}$.
The $1:2$ direct resonance has 
$a \sim h^{2/3}$ and the $1:1$ direct resonance is the weakest:
$a \sim h$. (The fact that the harmonic direct resonance is
weaker than the superharmonic one, is in agreement with
results of  \cite{QuinteroLetter,QuinteroDirect} where
it was established in the undamped situation.)

Our asymptotic approach also allows to rank the resonances according to 
the widths of the corresponding Arnold tongues
on the ``driving strength {\it vs\/}  driving frequency" plane.
The $1:1$ parametric resonance is the widest one; in this
case the resonant region is bounded by the curve $h \sim \rho^{3/2}$. 
The $2:1$ parametric resonance is second widest; in this case
the Arnold tongue has $h \sim \rho$. The $1:2$ direct resonance has 
$h \sim \rho^{3/4}$ and the $1:1$ direct resonance is the narrowest:
$h \sim \rho^{1/2}$. We should also mention that the $1:1$ parametric 
and the $1:2$ direct resonance have no threshold in the strength of the
driver
whereas the $2:1$ parametric and $1:1$ direct resonances occur
only if the driving strength exceeds a certain threshold value.

\begin{acknowledgments}
We thank 
Alan Champneys  for useful remarks.
O.O. was supported by funds provided by the National Research Foundation 
of South Africa
and the University of Cape Town. 
I.B. was supported by the NRF under
grant 2053723.
\end{acknowledgments}

\end{document}